\documentclass[prl,twocolumn]{revtex4-2}
\usepackage{mathrsfs}
\usepackage{amsmath}
\usepackage{amssymb}
\usepackage{graphicx}
\usepackage{braket}
\usepackage{hyperref}
\usepackage{color}
\usepackage{dsfont}
\usepackage{physics}
\usepackage{mathtools}

\newcommand{\orcid}[1]{\href{https://orcid.org/#1}{\includegraphics[height=\fontcharht\font`\B]{ORCIDiD_icon16x16.png}}}

\hypersetup{
    colorlinks,
    citecolor=blue,
    linkcolor=blue,
    urlcolor=blue
}

\begin{document}
\title{Quantum Walks with Indefinite Causal Order}
\author{Yuanbo Chen}
\email{chen@biom.t.u-tokyo.ac.jp}
\author{Yoshihiko Hasegawa}
\email{hasegawa@biom.t.u-tokyo.ac.jp}
\affiliation{Department of Information and Communication Engineering, Graduate
School of Information Science and Technology, The University of Tokyo,
Tokyo 113-8656, Japan}

\date{\today}
\begin{abstract}
In all existing quantum walk models, the assumption about a pre-existing fixed background causal structure is always made and has been taken for granted. Nevertheless, in this work we will get rid of this tacit assumption especially by introducing \emph{indefinite causal order} ways of coin tossing and investigate this modern scenario. We find that an ideal-shape and fast-spreading uniform distribution can be prepared with our new model. First we will show how an always-symmetrical instantaneous distribution appears in an indefinite causal order quantum walk, which then paves the way for deriving conditions that enables one to interpret an evolved state as a superposition of its \emph{definite} causal order counterparts which is however in general prohibited. This property of temporal superposition of states can be generalized from two-process scenarios to cases where arbitrary many $\mathcal{N}$ processes are involved. Finally, we demonstrate how a \emph{genuine} uniform distribution emerges from an indefinite causal order quantum walk. Remarkably, besides the ideal shape of the distribution, our protocol has another powerful advantage, that is the speed of spatial spreading reaches exactly the theoretical limit in contrast to conventional cases where one may encounter the well-known issue of a $1/\sqrt{2}$ degeneration.
\end{abstract}
\maketitle

\emph{Introduction.---} The laws of quantum mechanics has been the driving force behind a broad range of modern technologies \cite{Brunner:2014:RMP,Ekert:1991:PRL,Nielsen:2011:QCQI,Giovannetti:2006:2006,Horodecki:2009:RMP,Brunner:2014:RMP}. In quantum information science, as the product of quantizing classical random walk models, quantum walks \cite{Aharonov:1993:PRA} is a good example. Practical values such as in the implementation of various algorithms \cite{Qu:2022:PRL} have made it become an important research field. Quantum walks is also of significant interest for its role as platforms to simulate versatile physical phenomenon \cite{Wang:2019:PRL,Xiao:2017:natphys,Ramasesh:2017:PRL}.

In conventional models of quantum walk, for instance in some discrete-time walk models, the state of a walker lives in a position Hilbert space, with a quantum mechanical coin equipped with a Hilbert space of finite dimension. Besides, a \emph{fixed} background causal structure is also assumed, which naturally enforces all operations have to be performed in a \emph{definite} causal order way. However, this assumption may not necessarily be true and the investigation of new possibilities has been made recently \cite{Oreshkov:2012:NCOMMS,Chiribella:2013:PRA,Araujo:2014:PRL,Guerin:2016:PRL,Rubino:2021:PRR,Chen:2021:arXiv_a,Chen:2021:arXiv_b}, which trigged the finding of a novel class of quantum processes whereby one can witness the property of \emph{indefinite causal order} \cite{Rubino:2017:SCIADV,Goswami:2018:PRL,Procopio:2015:NCOMMS}.

In fact we have been doing the coin tossing in the same way as of a classical random walk, that is we toss the coin in a \emph{fixed causal order} manner. Ever since the revolution when random walk models experienced the transition from classical to quantum we have benefited a lot in fruitful ways. Nonetheless, the belief that promoting the operational aspects of a model would provide us with further advantages drives one to begin the exploration into new possibilities.

In this Letter, we present several findings which are not only interesting in its own right, but provide valuable insight into both the field of quantum walk and indefinite causal order. Especially, the results further deepened our knowledge of indefinite causal order by unravelling several new facts. First, we will show that regardless of the condition that a coin is \emph{not} prepared in a quantum superposition state, in an indefinite causal order quantum walk, one is \emph{always} guaranteed with a symmetrical instantaneous spatial distribution even for \emph{any conceivable} coin tossing rules. Second, we demonstrate that although generally an evolved state from indefinite causal order dynamics may not necessarily be considered as a result of superposing several definite causal order evolved states. However, providing that certain conditions are satisfied one is allowed to interpret such an evolved states as a superposition of whose definite causal order counterparts. Although we will first derive that property in the case where a \emph{2}$-$Switch is employed, it also holds with the  \emph{2}$-$Switch replaced by a cyclic $\mathcal{N}-$Switch.

We then find that with this elegant property of temporal superposition it becomes possible to fulfill the goal of preparing \emph{genuine} spatial uniform distributions. In addition to the \emph{perfect} (\emph{ideal}) shape of uniform distributions, the speed of spatial spread reaches exactly the theoretical limit in contrast to the degeneration characterized by a factor of $1/\sqrt{2}$ in many conventional definite causal order quantum walks. A distribution longer in length can be obtained via a cyclic $\mathcal{N}-$Switch with lager $\mathcal{N}$. It is both the temporal and spatial quantum superpositions give rise to the amazing finding.

\emph{Discrete time coin--walker model.---} Let us first introduce the mathematical description of the model and establish the notations. We consider a quantum walker moving along a line, whose position state is associated with a Hilbert space $\mathcal{H}^W$ with the basis $\{\ket{n}:n\in\mathbb{Z}\}$. The quantum walk model under our investigation throughout this work is restricted to be discrete, that is a quantum-mechanical coin will be used to indicate the direction towards which the walker proceeds. 

Depending on the state of the coin, the direction of the walker's movement may be different. Specifically, since we are dealing with one-dimensional quantum walks, the local Hilbert space of the coin $\mathcal{H}^C$ is then two-dimensional which is spanned by the basis $\{\ket{\vartriangleright},\ket{\blacktriangleleft}\}$. With these conventions, we define the displacement operator for the walker as
\begin{align}
	\label{eq:def_shifter}
	\hat{\mathcal{S}}
	&=\sum_{n\in\mathbb{Z}}\ketbra{\vartriangleright,{n+1}}{\vartriangleright,{n}}+\ketbra{\blacktriangleleft,{n-1}}{\blacktriangleleft,{n}}.
\end{align}
In Eq.~\eqref{eq:def_shifter} shifter $\hat{\mathcal{S}}\in{\mathcal{L}}({\mathcal{H}^C}{\otimes{\mathcal{H}^W}})$ conditionally moves the walker \emph{forward} one site if the coin state is $\ket{\vartriangleright}$, otherwise the walker will go \emph{backward} when the coin occupies the state $\ket{\blacktriangleleft}$, where $\mathcal{L}(\cdot)$ stands for the set of linear operators acting on a Hilbert space. In addition to the shifter, we also need coin operators $\hat{\mathcal{C}}\in{\mathcal{L}}({\mathcal{H}^C})$ to carry out coin tossing just like what one does in the classical random walk. 

Here, we make the choice of employing SU(2) coin operators, the most general coin operator for a two-dimensional coin, which constitute a family of three-parameter linear operators whose definition is
\begin{align}
	\label{eq:def_su2}\nonumber
	\hat{\mathcal{C}}(\alpha,\beta,\theta)=e^{i\alpha}\cos{\theta}&\ketbra{\vartriangleright}{\vartriangleright}+~~e^{i\beta}\sin{\theta}\ketbra{\vartriangleright}{\blacktriangleleft}\\
	+e^{-i\beta}\sin{\theta}&\ketbra{\blacktriangleleft}{\vartriangleright}-e^{-i\alpha}\cos{\theta}\ketbra{\blacktriangleleft}{\blacktriangleleft}.
\end{align}
\begin{figure*}[t]
    \centering
    \includegraphics[width=0.96\textwidth]{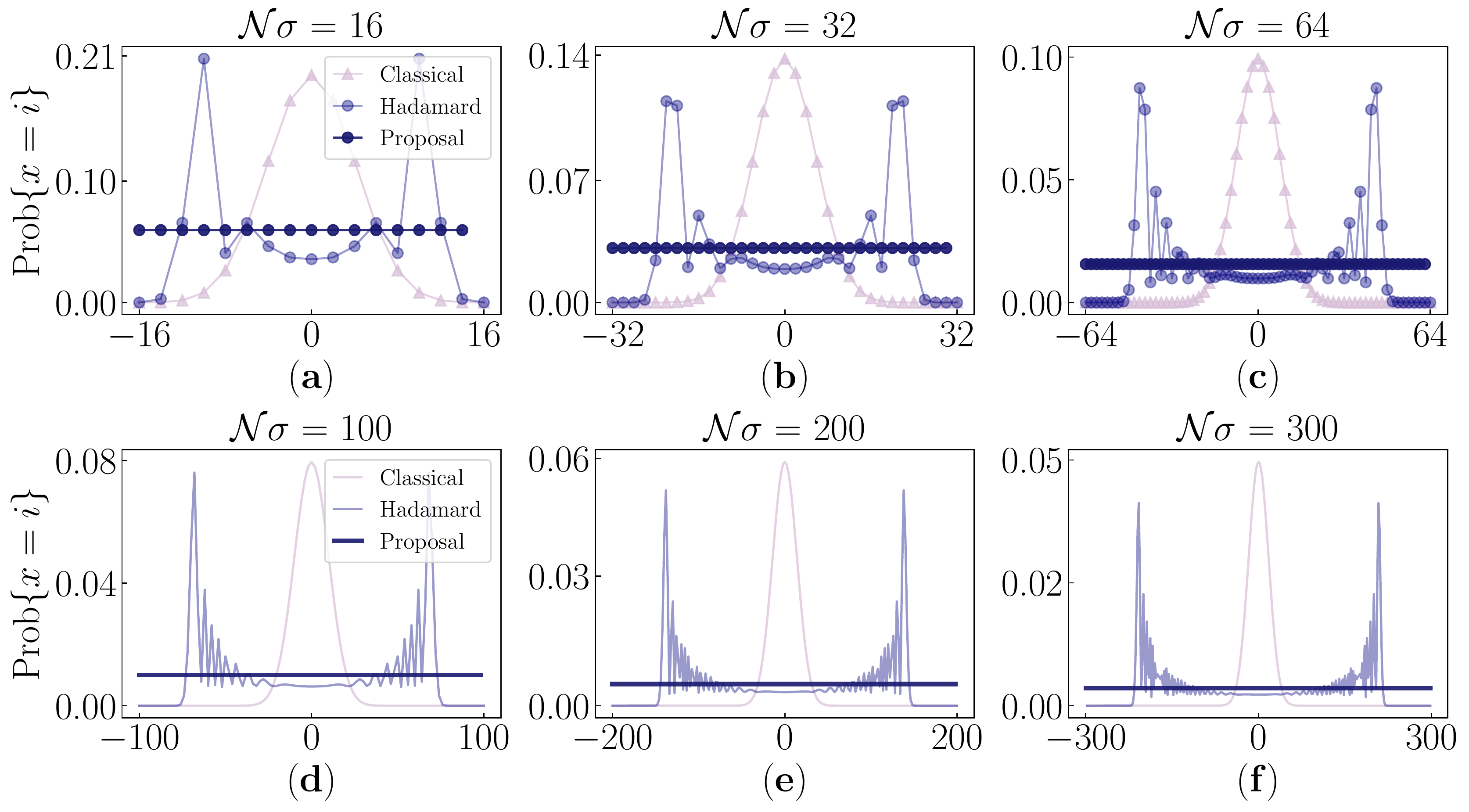}
    \caption{Intuitive illustration is presented here. We have plotted in (a) a classical random walk, a normal Hadamard walk and our proposal for comparison where the number of steps is taken to be 16. (a)---(f) differs in the steps such that we choose to observe how the behavior of a random walk varies from 16--step to 32, 64, 100, 200 and 300--step cases. It is clear from the figures that our uniform distribution always spreads at a speed of theoretical limit.}
    \label{fig:comparison}
\end{figure*}
The state of the walker plus the coin after $\sigma$ steps of time evolution is obtained by applying the time propagator $\sigma$ times, so that the corresponding unitary transformation reads $\hat{\mathbb{U}}^\sigma$, where $\hat{\mathbb{U}}$ is the single-step time propagator which is expressed as $\hat{\mathbb{U}}=\hat{\mathcal{S}}\circ(\hat{\mathcal{C}}\otimes\hat{\mathbb{I}}_W)$, with $\hat{\mathcal{C}}$ being an operator defined according to Eq.~\eqref{eq:def_su2}.

We turn now to study our novel scenarios where the causal order of coin tossing among distinct points may no longer follow a classical fashion. Before that let us first introduce two sets of coin operators $\{\hat{\mathcal{C}_i}^{p_0}\}_{i=0}^{\sigma_0-1}$ and $\{\hat{\mathcal{C}_i}^{p_1}\}_{i=0}^{\sigma_1-1}$, each characterizing a $\sigma_{0(1)}-$step state update rule. They are given the names $\hat{\mathbb{U}}^{\sigma_0}$ and $\hat{\mathbb{U}}^{\sigma_1}$ respectively. Because the index $i$ is merely a label, that is the arguments in $\hat{\mathcal{C}_i}^{p_0}=\hat{\mathcal{C}}(\alpha_i^{p_0},\beta_i^{p_0},\theta_i^{p_0})$ are in fact free to be any valid values, therefore here we simply define $\hat{\mathbb{U}}^{\sigma_{1}}$ to be
\begin{align}
	\label{eq:def_u_sigma_1}
	\hat{\mathbb{U}}^{\sigma_0}=\hat{\mathcal{U}}_{\sigma_0-1}^{p_0}\circ\hat{\mathcal{U}}_{\sigma_0-2}^{p_0}\circ\dots\circ\hat{\mathcal{U}}_{1}^{p_0}\circ\hat{\mathcal{U}}_{0}^{p_0},
\end{align}
where $\hat{\mathcal{U}}_{i}^{p_0}=\hat{\mathcal{S}}\circ(\hat{\mathcal{C}}_i^{p_0}\otimes\hat{\mathbb{I}}_W)$ and the same also applies to the definition of $\hat{\mathbb{U}}^{\sigma_1}$. One notices that \emph{all} possible $\sigma_0-$step state update rules are in fact encoded in Eq.~\eqref{eq:def_u_sigma_1} since we have not made any declaration about specific mapping rules between the parameter set $\{(\alpha_i^{p_0},\beta_i^{p_0},\theta_i^{p_0})\}_{i=0}^{\sigma_0-1}$ and $i$.

Conventionally, even in the traditional quantum mechanics setting, for the two operators $\hat{\mathbb{U}}^{\sigma_0}$ and $\hat{\mathbb{U}}^{\sigma_1}$, the causal order of which is trivial, i.e., the only two possibilities are either $\hat{\mathbb{U}}^{\sigma_0}$ takes place before $\hat{\mathbb{U}}^{\sigma_1}$ ($\hat{\mathbb{U}}^{\sigma_1}\circ\hat{\mathbb{U}}^{\sigma_0}$) or the reverse ($\hat{\mathbb{U}}^{\sigma_0}\circ\hat{\mathbb{U}}^{\sigma_1}$). However, it was not until recently the possibility of superposing time evolutions with different causal orders had been discussed, and in fact one can use a physical resource known as the quantum switch to realize this new class of processes.

\emph{Quantum walks without definite causal order.---}
A quantum \emph{2}$-$Switch has the following mathematical descrption,
\begin{align}
	\label{eq:def_qsw}\nonumber
	\hat{\mathbb{V}}^{}=&\ketbra{0}{0}^\mathcal{O}\otimes{\hat{\mathbb{U}}^{\sigma_1}}\circ{\hat{\mathbb{U}}^{\sigma_0}}\\
	+&\ketbra{1}{1}^\mathcal{O}\otimes{\hat{\mathbb{U}}^{\sigma_0}}\circ{\hat{\mathbb{U}}^{\sigma_1}}.
\end{align}
Basically, with a full quantum switch one is able of, in principle, superposing all permutations of a given number of processes. However, it should be noticed that treating a full permutation of, say, a 100-step walk is daunting, and experimentally prohibitive at least within the foreseeable future. Here, we choose to wrap $\sigma-$step evolution into a single unitary propagator.

We are now ready to do calculations to investigate the dynamics of quantum walks under indefinite causal order scenario. Our walker starts at a spatially localized state, let us call it $\ket{n_0}$, with the coin initialized in the state $\ket{\phi_0}$, where $\phi_0\in\{\vartriangleright,\blacktriangleleft\}$. For the order control system, we prepare it in a balanced state, such that $\ket{\psi_0}^{\mathcal{O}}=(\ket{0}^{\mathcal{O}}+\ket{1}^{\mathcal{O}})/\sqrt{2}$. The composite system then evolves under the action of Eq.~\eqref{eq:def_qsw} in the way that ${\rho}^{\mathcal{O}CW}=\hat{\mathbb{V}}^{}{\rho}_{\star}^{\mathcal{O}CW}\hat{\mathbb{V}}^{\dagger}$. Here ${\rho}_{\star}^{\mathcal{O}CW}\coloneqq{\rho}_{\star}^{\mathcal{O}}\otimes{\rho}_{\star}^{C}\otimes{\rho}_{\star}^{W}$, where ${\rho}_{\star}^{\mathcal{O}}\coloneqq\ketbra{\psi_0}{\psi_0}^{\mathcal{O}}$, ${\rho}_{\star}^{C}\coloneqq\ketbra{\phi_0}{\phi_0}^{C}$ and ${\rho}_{\star}^{W}\coloneqq\ketbra{n_0}{n_0}^{W}$, with the subscript $\ast$ highlighting an initial state. Following the time evolution, a measurement will be performed on the order control system in the Fourier basis $\big\{\ket{\mathcal{F}_m^d}\big\}_{m=0}^{d-1}$ (here, we take $d=2$) defined as follows,
\begin{align}
	\label{eq:def_fourier}
	\ket{\mathcal{F}_m^d}=\frac{1}{\sqrt{d}}\sum_ke^{-i\frac{2\pi m}{d}k}\ket{k}^{\mathcal{O}},
\end{align}
where $d$ is the dimension of the order control system, with $m$ denoting the index of elements in the measurement basis. Therefore we have projectors $\hat{\mathcal{P}}_{\mathcal{F}_m^d}\coloneqq\ketbra{\mathcal{F}_m^d}{\mathcal{F}_m^d}$, and depending on the measurement outcome, $\rho^{\mathcal{O}CW}$ collapses into different branches of future. Let us denote by $\rho_{|\mathcal{F}_i^2}^{CW}$ all the possible outcomes resulting from a measurement made over the order control system. By a straightforward calculation, we can express $\rho_{|\mathcal{F}_i^2}^{CW}$ as
\begin{align}
	\label{eq:rho_cw_fi2}\nonumber
	\rho_{|\mathcal{F}_i^2}^{CW}&={\rm{Tr}}_{\mathcal{O}}\big[\hat{\mathcal{P}}_{\mathcal{F}_i^2}\rho^{\mathcal{O}CW}\hat{\mathcal{P}}_{\mathcal{F}_i^2}\big]\\\nonumber
	&=\sum_j\big \langle j\ketbra{\mathcal{F}_i^2}{\mathcal{F}_i^2}^{\mathcal{O}}\rho^{\mathcal{O}CW}\ketbra{\mathcal{F}_i^2}{\mathcal{F}_i^2}j\big \rangle^{\mathcal{O}}\\
	&=\bra{\mathcal{F}_i^2}\rho^{\mathcal{O}CW}\ket{\mathcal{F}_i^2}.
\end{align}
In Eq.~\eqref{eq:rho_cw_fi2}, it should be noticed that to have a physical density operator one should further do the normalization, i.e., $\rho_{|\mathcal{F}_i^2}^{CW}/{\rm{Tr}}\big[\rho_{|\mathcal{F}_i^2}^{CW}\big]$. We will omit this procedure also for other density operators due to the reason that will become clear later in this manuscript.

\emph{Always symmetrical instantaneous distributions.---} The spatial distribution of an evolved walker is especially of interest and here we restrict our attention on the \emph{instantaneous} spatial distribution. In order to do this, we further split $\rho_{|\mathcal{F}_i^2}^{CW}$ into two conditional states which are the projected states corresponding to results induced by measurements on the coin with projectors $\{\ketbra{\vartriangleright}{\vartriangleright},\ketbra{\blacktriangleleft}{\blacktriangleleft}\}$. We use the symbols $\rho_{|\phi,\mathcal{F}_i^2}^{W}$ to represent them. The corresponding spatial probability distribution is encoded in the population terms of a density operator represented in terms of the position basis. In our language, it reads ${\rm{Prob}}\{x=i\}=\bra{i}\rho\ket{i}$. We now show an astounding finding about $\rho_{|\phi,\mathcal{F}_i^2}^{W}$ whereby the population terms display a symmetry property.

\textbf{Theorem 1:} Let $\{\hat{\mathcal{C}_i}^{p_0}\}_{i=0}^{\sigma_0-1}$ and $\{\hat{\mathcal{C}_i}^{p_0}\}_{i=0}^{\sigma_1-1}$ be two sets of arbitrary different coin operators, where $\hat{\mathcal{C}_i}^{p_0(p_1)}\in\rm{SU(2)}$ and $\sigma_{0(1)}\geq1$. Consider that a walker starts from the state $\ketbra{n_0}{n_0}$ with the coin initialized in $\ketbra{\phi_0}{\phi_0}$, where $\phi_0\in\{\vartriangleright,\blacktriangleleft\}$. The walker undergoes indefinite causal order walk dynamics where the coin tossing rule is subject to $\hat{\mathcal{C}_i}^{{p_0}({p_1})}$. Then the following always holds,
\begin{align}
	\label{eq:theorem_1}\nonumber
	\bra{n_0-l}\rho_{|\phi_0,\mathcal{F}_1^2}^W\ket{n_0-l}=\bra{n_0+l}\rho_{|\phi_0,\mathcal{F}_1^2}^W\ket{n_0+l},\\
	{\rm{for~}}\forall~n_0,l\in\mathbb{Z}.
\end{align}

\emph{Proof.---} The proof is in Sec. S1 of the Supplementary Material.

Theorem 1 implies that whenever the condition described above is satisfied, a walker's spatial probability distribution is symmetrical with respect to the place from where it starts. This behavior is far from being contrary to our common sense if one notices, firstly, that the coin is \emph{not} in a quantum superposition state. Second, the coin operators can be chosen freely to any degree. 

Indeed, to witness a symmetrical instantaneous probability distribution, one should carefully prepare the coin in a quantum superposition state, however this is far from enough. In addition to the initial state of the coin, the tossing rule must be designed depending on the coin's initial state.

By virtue of Theorem 1, of course one is always guaranteed with a symmetrical probability distribution, and then it is up to oneself to design the shape of the distribution by choosing the coin operators. However, why not first consider an extreme case? That is the most trivial symmetrical distribution --- the zero probability distribution. We want to establish conditions to make this possible. In what follows, we will provide such a sufficient condition which is non-trivial and prove it really results in a zero probability distribution based on Lemma 1 which we now introduce.

At this point we shall remind one of a notational convention that any quantum state written by a $\rho$ is associated with an indefinite causal order dynamics, whereas the symbol $\varrho$ is reserved for representing a density operator obtained by evolving the system with \emph{definite} causal order dynamics.

\textbf{Lemma 1:} Let $\varrho_{0|\phi}^W$ and $\varrho_{1|\phi}^W$ denote the conditional states obtained by applying $\{\hat{\mathbb{U}}^{\sigma_0},\hat{\mathbb{U}}^{\sigma_1}\}$ in the order of $\hat{\mathbb{U}}^{\sigma_1}\circ\hat{\mathbb{U}}^{\sigma_0}$ and $\hat{\mathbb{U}}^{\sigma_0}\circ\hat{\mathbb{U}}^{\sigma_1}$ respectively, where $\phi\in\{\vartriangleright,\blacktriangleleft\}$, then we have the following relation,
\begin{align}
	\label{eq:lemma_1}
	\rho_{|\phi,\mathcal{F}_0^2}^W+\rho_{|\phi,\mathcal{F}_{1}^2}^W=\frac{1}{2}(\varrho_{0|\phi}^W+\varrho_{1|\phi}^W).
\end{align}

\emph{Proof.---} The proof is in Sec. S2 of the Supplementary Material.

We now show the above-mentioned sufficient condition and proof below.

\textbf{Corollary 1:} Let $\{\hat{\mathcal{C}_i}^{p_0}\}_{i=0}^{\sigma_0-1}$ be a set of arbitrary different coin operators while $\{\hat{\mathcal{D}_i}^{p_1}\}_{i=0}^{\sigma_1-1}$ is another set of coin operators whose definition takes the form $\hat{\mathcal{D}}_i^{p_1}=\hat{\mathcal{C}}(\alpha_i^{p_1},\beta_i^{p_1},0)$ where $\sigma_{1}\geq\sigma_{0}\geq1$. Consider that the coin is initialized to be $\ketbra{\phi_0}{\phi_0}$, where $\phi_0\in\{\vartriangleright,\blacktriangleleft\}$. The walker undergoes indefinite causal order walk dynamics, then there is no ambiguity in determining $\rho_{|\phi_0,\mathcal{F}_1^2}^W$, or $\rho_{|\phi_0,\mathcal{F}_1^2}^W=0$.

\emph{Proof.---} The proof is in Sec. S3 of the Supplementary Material.

One should notice that even under situations where the instantaneous probability distributions for each chirality do not vanish in definite causal order scenarios, Corollary 1 is always true. Consequently, this uncertainty before a measurement is made on the coin has been completely eliminated, i.e., $\rho_{\mathcal{F}_1^2}^{W}=\rho_{|\phi,\mathcal{F}_1^2}^{W}$.

The uncertainty originating from the measurements is basically at the maximal degree, in other words, the four possible reduced states $\rho_{|\phi,\mathcal{F}_i^2}^W$ are different from each other. We have already reduced the number of four to three by making one member vanishes. It turns out that a further step can be made to reduce this uncertainty --- we are not at the end of the story.

\textbf{Lemma 2:} Under the condition described in Corollary 1, two possible branches of measurement outcomes become equivalent for their population terms, that is $\rho_{|\phi^{\bot}_0,\mathcal{F}_0^2}^W \circeq \rho_{|\phi^{\bot}_0,\mathcal{F}_1^2}^W$, and for the coherence terms, they are identical up to some phase differences. Note that hereafter we will use $\circeq$ to represent such a kind of equivalence relation.

\emph{Proof.---} The proof is in Sec. S4 of the Supplementary Material.

With Lemma 2 in our hands, one is allowed to write $\rho_{|\phi^{\bot}_0,\mathcal{F}_i^2}^W$ as a \emph{superposition} of two evolved states related to classical causal order dynamics.

\textbf{Corollary 2:} Following Lemma 2, we find that $\rho_{|\phi^{\bot}_0,\mathcal{F}_i^2}^W$ is a linear combination of two states resulting from classical causal order dynamics, it takes the form
\begin{align}
	\label{eq:corollary_2}
	\rho_{|\phi^{\bot}_0,\mathcal{F}_i^2}^W \circeq \frac{1}{4}(\varrho_{0|\phi^{\bot}_0}^W+\varrho_{1|\phi^{\bot}_0}^W).
\end{align}

\emph{Proof.---} The proof is in Sec. S5 of the Supplementary Material.

\textbf{Lemma 3:} For $\varrho_{0|\phi^{\bot}_0}^W$ and $\varrho_{1|\phi^{\bot}_0}^W$ in Eq.~\eqref{eq:corollary_2}, they are identical up to a translation in the position direction by a length of $\sigma_1$, or $\bra{x}\varrho_{1|\phi^{\bot}_0}^W\ket{x}=\bra{x-2\sigma_1}\varrho_{0|\phi^{\bot}_0}^W\ket{x-2\sigma_1}$, for $\forall~x,y\in\mathbb{Z}$.

\emph{Proof.---} The proof is in Sec. S6 of the Supplementary Material.

Eq.~\eqref{eq:corollary_2} gives us the implication that we are now able to skip direct calculations to reach the goal. Moreover, Lemma 3 tells us that it suffices to know $\varrho_{0|\phi^{\bot}_0}^W$ or $\varrho_{1|\phi^{\bot}_0}^W$ only, one is then naturally led to raise the question of how to benefit from Corollary 2. We now begin to demonstrate such an application --- how \emph{the~genuine} uniform probability distribution appears from an indefinite causal order quantum walk.

Let us consider a special case where $\ketbra{n_0}{n_0}=\ketbra{0}{0}$ and $\{\hat{\mathcal{C}}_i^{p_0}\}_{i=0}^{\sigma_0-1}$ are all identical and equal to $\hat{\mathcal{C}}(0, 0,\pi/4)$ with $\sigma_i=2$. One finds that the distribution corresponding to $\varrho_{0|\phi^{\bot}_0}^W/{\rm{Tr}}\big[\varrho_{0|\phi^{\bot}_0}^W\big]$ is Prob$\{x=i\}=1/2$, for $i\in\{0,2\}$.  So we quickly conclude that the distribution for $(\rho_{|\phi^{\bot}_0,\mathcal{F}_0^2}^W+\rho_{|\phi^{\bot}_0,\mathcal{F}_1^2}^W)/{\rm{Tr}}\big[\rho_{|\phi^{\bot}_0,\mathcal{F}_0^2}^W+\rho_{|\phi^{\bot}_0,\mathcal{F}_1^2}^W\big]$ is Prob$\{x=i\}=1/4$, for $i\in\{-4,-2,0,2\}$. One finds that the choice of $\sigma_i=4$ also gives us a uniform distribution with the length of eight steps.

By saying \emph{genuine}, the statement is endowed with two folds of meaning. First, our uniform distribution is by no means an artefact which is in reality merely an approximation to our result, or, \emph{the ideal} ones. Secondly, since the uncertainty during the coin measurement is completely removed, there is no need for one to worry about to which chirality the distribution is associated with. In other words, unlike previous studies on this topic whereby the approximated uniform-like distribution is the sum of the two skewed distribution for each chirality, with our method however, as long as the desirable order control system measurement results appears, we are at the goal.

\emph{Genuine uniform distribution for arbitrary long even length steps.---} The preceding discussion is limited to the \emph{2}$-$Switch case, and the length of the uniform distribution can not go beyond eight. In the following we are going to generalize the already established result so that one can have a distribution which spreads farther. This goal is made by considering indefinite causal order processes involving more than two subprocesses. 

We begin by first introducing $\mathcal{N}$ sets of SU(2) coin operators, and further divide them into two groups. They are denoted by $\{\hat{\mathcal{C}}_i^{p_0}\}_{i=0}^{\sigma_0-1}$ and $\{\hat{\mathcal{D}}_i^{p_n}\}_{i=0}^{\sigma_n-1}$ respectively, where $n$ runs from $1$ to $\mathcal{N}-1$. For $\hat{\mathcal{C}}_i^{p_0}$, we assume that they can be arbitrary SU(2) coin operators whereas the third parameter argument of all $\hat{\mathcal{D}}_i^{p_n}$ is fixed to be zero, that is
\begin{align}
	\label{eq:def_coin_cd}\nonumber
	\hat{\mathcal{C}}_i^{p_0}&=\hat{\mathcal{C}}(\alpha_i,\beta_i,\theta_i),\\
	\hat{\mathcal{D}}_i^{p_n}&=\hat{\mathcal{C}}(\alpha_i^{p_n},\beta_i^{p_n},0).
\end{align}

Inspired by the results obtained so far, we shall use a \emph{cyclic} $\mathcal{N}-$Switch to increase the number of subprocess, the action of which on the total system is mathematically described by
\begin{align}
	\label{eq:def_qcnsw}
	\hat{\mathbb{W}}&_{\mathcal{N}}^{}=\sum_{n=0}^{\mathcal{N}-1}\ketbra{n}{n}^\mathcal{O}\otimes\mathbb{L}_n\big\{{\hat{\mathbb{U}}^{\sigma_{\mathcal{N}-1}}}\circ\dots\circ{\hat{\mathbb{U}}^{\sigma_0}}\big\}.
\end{align}
where we have used the symbol $\mathbb{L}_n\big\{\dots\big\}$ to denote the $n-$th left cyclically shifted composition of a string of operators, e.g., the right-most operator in the bracket is shifted to the left $n$ sites, and the same for other operators.

Let us first show a generalization of Lemma 2 and Corollary 2.

\textbf{Theorem 2:} In a quantum walk described by Eq.~\eqref{eq:def_coin_cd} and Eq.~\eqref{eq:def_qcnsw} where the walker starts from an arbitrary spatially localized state $\ketbra{n_0}{n_0}$ with the coin initialized in $\ketbra{\phi_0}{\phi_0}$ and $\ket{\psi_0}^{\mathcal{O}}=\sum_{j=0}^{\mathcal{N}-1}\ket{j}^{\mathcal{O}}/\sqrt{\mathcal{N}}$, where $n_0\in\mathbb{Z}$ and $\phi_0\in\{\vartriangleright,\blacktriangleleft\}$. Providing that $\sigma_i\geq\sigma_0$, for $\forall~i\neq0$, we have $\rho_{|\phi^{\bot}_0,\mathcal{F}_m^\mathcal{N}}^W \circeq \rho_{|\phi^{\bot}_0,\mathcal{F}_n^\mathcal{N}}^W$, for $\forall~m\neq n \in[0,\mathcal{N}-1]$, with the following relation being true
\begin{align}
	\label{eq:theorem_2}
	\rho_{|\phi^{\bot}_0,\mathcal{F}_i^{\mathcal{N}}}^W \circeq \frac{1}{\mathcal{N}^2}\sum_{n=0}^{\mathcal{N}-1}\varrho_{n|\phi^{\bot}_0}^W,
\end{align}
where $\varrho_{n|\phi^{\bot}_0}^W$ means that it corresponds to a time evolved state propagated by $\mathbb{L}_n\big\{{\hat{\mathbb{U}}^{\sigma_{\mathcal{N}-1}}}\circ\dots\circ{\hat{\mathbb{U}}^{\sigma_0}}\big\}$.

\emph{Proof.---} The proof is in Sec. S7 of the Supplementary Material.

Our protocol for creating a uniform distribution is of a try-until-success fashion, due to this feature one must take the probability issue into account.

\textbf{Theorem 3:} Consider the walker's initial state to be $\ketbra{n_0}{n_0}$, and for simplicity take $\sigma_i$ to be $2$ or $4$, with the the condition in Theorem 2. We choose $\hat{\mathcal{C}}_i^{p_0}=\hat{\mathcal{C}}(0,0,\pi/4)$. The distribution associated with the states $\rho_{|\phi^{\bot}_0,\mathcal{F}_i^{\mathcal{N}}}^W$ resulting from a cyclic $\mathcal{N}-$ Switch is uniform, or
\begin{align}
	\label{eq:theorem_3}\nonumber
	{\rm{Prob}}\{x=i+n_0&\}=\frac{1}{\mathcal{N}\sigma_i},\\
	{\rm{for}}~i\in[-\mathcal{N}\sigma_i,~&\mathcal{N}\sigma_i-2],
\end{align}
where $i$ is an even integer. The probability of success is 1/2 for the case of $\sigma_i=2$ and 1/4 for the case of $\sigma_i=4$, where $\mathcal{N}\geq2$.

\emph{Proof.---} The proof is in Sec. S8 of the Supplementary Material.

The shape of our uniform distribution is shown in Fig~\ref{fig:comparison}, and the comparison with two conventional kinds of random walk can be observed easily.

In fact, although there are in principle $2\times\mathcal{N}$ possible distinct measurement outcomes, this ambiguity is drastically reduced to two. To clarify this, we display all the $2\times\mathcal{N}$ outcomes, among which $\mathcal{N}$ are identical reduced states as a consequence of Theorem 2, a special one is the $\rho_{\phi_0|\mathcal{F}_0^{\mathcal{N}}}^W$, and the remaining $\mathcal{N}-1$ possible ones come with zero probability. Therefore $2\times\mathcal{N}$ possible outcomes fall into just two groups which is the same as what we have seen before.

It should be noticed that the distribution Eq.~\eqref{eq:theorem_3} is not only of the ideal shape but the spatial propagation is at the $fastest$ possible rate. Recall that in a definite causal order Hadamard walk, the walker's distribution propagates at the speed of $\mathcal{N}\sigma/\sqrt{2}$ as a consequence of the exponentially decreasing probability out of the region $[-\mathcal{N}\sigma/\sqrt{2},~\mathcal{N}\sigma/\sqrt{2}]$. Nevertheless, our result suggests that it is possible to make the spatial propagation rate reach the theoretical limit ($\mathcal{N}\sigma$) while retaining a perfect uniform shape.

 Let us now give some interpretations to our results. In general a state like $\rho_{|\phi^{\bot}_0,\mathcal{F}_i^{\mathcal{N}}}^W$ can \emph{not} be a linear combination of $\varrho_{n|\phi^{\bot}_0}^W$ because although we are superposing the time evolutions with different casual orders, the statement that a resulting reduced state can therefore be written in the form of an expansion in terms of, say, a \emph{definite causal order reduced state} basis, may not necessarily be true. Even so, what behind Theorem 2 is that in some cases we can actually interpret an \emph{indefinite causal order} reduced state as a quantum superposition of \emph{classical casual order} counterparts. It is this beautiful property enables us to make a fastest propagating perfect uniform distribution become possible.
 
 First, a different causal order of $\hat{\mathcal{D}}_i^{p_n}$ helps us propagate the walker to a different temporal location in the causal future with respect to other causal order combinations. Secondly, after the walker is ready at the start line in $\mathcal{N}$ causal futures, the $\hat{\mathcal{C}}_i^{p_0}$ spreads it out spatially. Finally, thanks to the temporal superposition (Theorem 2), all the $\mathcal{N}$ causal past-future path takes place simultaneously which then make all the $\mathcal{N}$ pieces exactly connected together to form a perfect uniform distribution. With the $\hat{\mathcal{C}}_i^{p_0}$ only, the best we can do is to get a four--step long uniform distribution. On the other hand, although $\hat{\mathcal{D}}_i^{p_n}$ admits the fastest possible spatial spreading, it results in a spatially localized state. Let us stress that Theorem 3 is exactly the combination of both these two advantages without any degeneration.

\emph{Conclusions.---} In the present work, we have shown that by introducing quantum uncertainty into the causal order of operations for quantum walk models, the behavior of a walker displayed different features in comparison with conventional models. Quite contrary to one's intuition, in cases of a \emph{2}$-$Switch being used, the spatial distribution of one branch always comes with a symmetrical shape. Taking this as a starting point, we further derived a sufficient condition under which an evolved state from indefinite causal order dynamics can be interpreted as a superposition of its classical scenario counterparts.

As an application using the property of temporal superposition of states, which we further generalized for $\mathcal{N}$ processes in Theorem 2, a protocol for preparing genuine uniform distributions is proposed. In our proposal, we demonstrated a surprising finding that our distribution is not only of the ideal shape, but it propagates at the fastest possible speed. In addition, the probability of success is also analyzed, which we find to have a universal value of 1/2 for $2\times\mathcal{N}-$step long distributions and 1/4 where the length of steps is $4\times\mathcal{N}$. To the best of our knowledge, this is the first work that achieves perfectly the theoretical limit for both of the two aspects at the same time.

\end{document}


\title{Supplementary Material for \\``Quantum Walks with Indefinite Causal Order''}
\author{Yuanbo Chen}
\email{chen@biom.t.u-tokyo.ac.jp}
\author{Yoshihiko Hasegawa}
\email{hasegawa@biom.t.u-tokyo.ac.jp}
\affiliation{Department of Information and Communication Engineering, Graduate
School of Information Science and Technology, The University of Tokyo,
Tokyo 113-8656, Japan}

\maketitle
This supplementary material describes the calculations introduced in the main text. Equation and figure numbers are prefixed with S (e.g., Eq.~(S1) or Fig.~S1). Numbers without this prefix (e.g., Eq.~(1) or Fig.~1) refer to items in the main text.

\section{Proof of Theorem 1}
Let us first explicitly write out the jointly evolved state of ${\rho}_{\star}^{\mathcal{O}CW}$ in the main text,
\begin{align}
	\label{eq:rho_ocw_derv}\nonumber
	{\rho}^{\mathcal{O}CW}=&\hat{\mathbb{V}}^{}{\rho}_{\star}^{\mathcal{O}CW}\hat{\mathbb{V}}^{\dagger}\\\nonumber
	=&(\ketbra{0}{0}^\mathcal{O}\otimes{\hat{\mathbb{U}}^{\sigma_1}}\circ{\hat{\mathbb{U}}^{\sigma_0}}+\ketbra{1}{1}^\mathcal{O}\otimes{\hat{\mathbb{U}}^{\sigma_0}}\circ{\hat{\mathbb{U}}^{\sigma_1}}){\rho}_{\star}^{\mathcal{O}}\otimes{\rho}_{\star}^{CW}\\
	&(\ketbra{0}{0}^\mathcal{O}\otimes{\hat{\mathbb{U}}^{\sigma_0 \dagger}}
\circ{\hat{\mathbb{U}}^{\sigma_1 \dagger}}+\ketbra{1}{1}^\mathcal{O}\otimes{\hat{\mathbb{U}}^{\sigma_1 \dagger}}\circ{\hat{\mathbb{U}}^{\sigma_0 \dagger}}).
\end{align}

Substituting Eq.~\eqref{eq:rho_ocw_derv} into Eq.~(6), we obtain
\begin{align}
	\label{eq:rho_cw_f02_derv}\nonumber
	\rho_{|\mathcal{F}_0^2}^{CW}=&\bra{\mathcal{F}_0^2}\rho^{\mathcal{O}CW}\ket{\mathcal{F}_0^2}\\\nonumber
	=&\frac{1}{\sqrt{2}}(\bra{0}^\mathcal{O}+\bra{1}^\mathcal{O})\rho^{\mathcal{O}CW}\frac{1}{\sqrt{2}}(\ket{0}^\mathcal{O}+\ket{1}^\mathcal{O})\\\nonumber
	=&\frac{1}{2}(\bra{0}^\mathcal{O}+\bra{1}^\mathcal{O})(\ketbra{0}{0}^\mathcal{O}\otimes{\hat{\mathbb{U}}^{\sigma_1}}\circ{\hat{\mathbb{U}}^{\sigma_0}}+\ketbra{1}{1}^\mathcal{O}\otimes{\hat{\mathbb{U}}^{\sigma_0}}\circ{\hat{\mathbb{U}}^{\sigma_1}})\\\nonumber
	&\Big[\frac{1}{2}(\ketbra{0}{0}^{\mathcal{O}}+\ketbra{0}{1}^{\mathcal{O}}+\ketbra{1}{0}^{\mathcal{O}}+\ketbra{1}{1}^{\mathcal{O}})\otimes{\rho}_{\star}^{CW}\Big]\\\nonumber
	&(\ketbra{0}{0}^\mathcal{O}\otimes{\hat{\mathbb{U}}^{\sigma_0 \dagger}}\circ{\hat{\mathbb{U}}^{\sigma_1 \dagger}}+\ketbra{1}{1}^\mathcal{O}\otimes{\hat{\mathbb{U}}^{\sigma_1 \dagger}}\circ{\hat{\mathbb{U}}^{\sigma_0 \dagger}})(\ket{0}^\mathcal{O}+\ket{1}^\mathcal{O})\\\nonumber
	=&\frac{1}{4}(\bra{0}^\mathcal{O}\otimes{\hat{\mathbb{U}}^{\sigma_1}}\circ{\hat{\mathbb{U}}^{\sigma_0}}+\bra{1}^\mathcal{O}\otimes{\hat{\mathbb{U}}^{\sigma_0}}\circ{\hat{\mathbb{U}}^{\sigma_1}})\\\nonumber
	&\Big[(\ketbra{0}{0}^{\mathcal{O}}+\ketbra{0}{1}^{\mathcal{O}}+\ketbra{1}{0}^{\mathcal{O}}+\ketbra{1}{1}^{\mathcal{O}})\otimes{\rho}_{\star}^{CW}\Big]\\\nonumber
	&(\ket{0}^\mathcal{O}\otimes{\hat{\mathbb{U}}^{\sigma_0 \dagger}}\circ{\hat{\mathbb{U}}^{\sigma_1 \dagger}}+\ket{1}^\mathcal{O}\otimes{\hat{\mathbb{U}}^{\sigma_1 \dagger}}\circ{\hat{\mathbb{U}}^{\sigma_0 \dagger}})\\\nonumber
	=&\frac{1}{4}({\hat{\mathbb{U}}^{\sigma_1}}\circ{\hat{\mathbb{U}}^{\sigma_0}}{\rho}_{\star}^{CW}{\hat{\mathbb{U}}^{\sigma_0 \dagger}}\circ{\hat{\mathbb{U}}^{\sigma_1 \dagger}}+{\hat{\mathbb{U}}^{\sigma_1}}\circ{\hat{\mathbb{U}}^{\sigma_0}}{\rho}_{\star}^{CW}{\hat{\mathbb{U}}^{\sigma_1 \dagger}}\circ{\hat{\mathbb{U}}^{\sigma_0 \dagger}}\\
	&~~~{\hat{\mathbb{U}}^{\sigma_0}}\circ{\hat{\mathbb{U}}^{\sigma_1}}{\rho}_{\star}^{CW}{\hat{\mathbb{U}}^{\sigma_0 \dagger}}\circ{\hat{\mathbb{U}}^{\sigma_1 \dagger}}+{\hat{\mathbb{U}}^{\sigma_0}}\circ{\hat{\mathbb{U}}^{\sigma_1}}{\rho}_{\star}^{CW}{\hat{\mathbb{U}}^{\sigma_1 \dagger}}\circ{\hat{\mathbb{U}}^{\sigma_0 \dagger}}).
\end{align}

Similarly, we also have
\begin{align}
	\label{eq:rho_cw_f12_derv}\nonumber
	\rho_{|\mathcal{F}_1^2}^{CW}=&\bra{\mathcal{F}_1^2}\rho^{\mathcal{O}CW}\ket{\mathcal{F}_1^2}\\\nonumber
	=&\frac{1}{\sqrt{2}}(\bra{0}^\mathcal{O}-\bra{1}^\mathcal{O})\rho^{\mathcal{O}CW}\frac{1}{\sqrt{2}}(\ket{0}^\mathcal{O}-\ket{1}^\mathcal{O})\\\nonumber
	=&\frac{1}{4}({\hat{\mathbb{U}}^{\sigma_1}}\circ{\hat{\mathbb{U}}^{\sigma_0}}{\rho}_{\star}^{CW}{\hat{\mathbb{U}}^{\sigma_0 \dagger}}\circ{\hat{\mathbb{U}}^{\sigma_1 \dagger}}-{\hat{\mathbb{U}}^{\sigma_1}}\circ{\hat{\mathbb{U}}^{\sigma_0}}{\rho}_{\star}^{CW}{\hat{\mathbb{U}}^{\sigma_1 \dagger}}\circ{\hat{\mathbb{U}}^{\sigma_0 \dagger}}\\
	&~{-\hat{\mathbb{U}}^{\sigma_0}}\circ{\hat{\mathbb{U}}^{\sigma_1}}{\rho}_{\star}^{CW}{\hat{\mathbb{U}}^{\sigma_0 \dagger}}\circ{\hat{\mathbb{U}}^{\sigma_1 \dagger}}+{\hat{\mathbb{U}}^{\sigma_0}}\circ{\hat{\mathbb{U}}^{\sigma_1}}{\rho}_{\star}^{CW}{\hat{\mathbb{U}}^{\sigma_1 \dagger}}\circ{\hat{\mathbb{U}}^{\sigma_0 \dagger}}).
\end{align}

Without loss of generality, we assume the initial state of the walker to be $\ketbra{0}{0}$ and the coin is initialized to occupy the state $\ketbra{{\vartriangleright}}{{\vartriangleright}}$. Following the definition of $\hat{\mathbb{U}}^{\sigma_{0(1)}}$ in the main text, we write the two unitary operators as
\begin{align}
	\label{eq:u_sigma01_explicit}
	\hat{\mathbb{U}}^{\sigma_{0(1)}}=\big[\hat{\mathcal{S}}\circ(\hat{\mathcal{C}}_{\sigma_{0(1)-1}}^{p_{0(1)}}\otimes\hat{\mathbb{I}}_W)\big]\circ\big[\hat{\mathcal{S}}\circ(\hat{\mathcal{C}}_{\sigma_{0(1)-2}}^{p_{0(1)}}\otimes\hat{\mathbb{I}}_W)\big]\circ\dots\circ\big[\hat{\mathcal{S}}\circ(\hat{\mathcal{C}}_{0}^{p_{0(1)}}\otimes\hat{\mathbb{I}}_W)\big].
\end{align}

Before preceding, let us introduce a parameterized operator which will turn out to be convenient later. Recall the definition of the shifter in the main text, Eq.~(1). We rewrite it in the form
\begin{align}
	\label{eq:def_shifter_another}
		\hat{\mathcal{S}}&=\sum_n\ketbra{\vartriangleright}{\vartriangleright}\otimes{\ketbra{n+1}{n}}+\ketbra{\blacktriangleleft}{\blacktriangleleft}\otimes{\ketbra{n-1}{n}}.
\end{align}
Let us now introduce the operator $\mathcal{X}_a$, which has the following definition
\begin{align}
	\label{eq:def_opx}
	\mathcal{X}_a\coloneqq\sum_n{\ketbra{n+a}{n}},
\end{align}
where, $a\in\mathbb{Z}$. Then Eq.~\eqref{eq:def_shifter_another} becomes
\begin{align}
	\label{eq:def_shifter_with_x}
	\hat{\mathcal{S}}&=\ketbra{\vartriangleright}{\vartriangleright}\otimes\mathcal{X}_1+\ketbra{\blacktriangleleft}{\blacktriangleleft}\otimes\mathcal{X}_{-1}.
\end{align}
Intuitively, $\mathcal{X}_a$ and $\mathcal{X}_b$ satisfies the the following algebra for $\forall~a,b\in\mathbb{Z}$,
\begin{align}
	\label{eq:xab_algebra_0}
	\mathcal{X}_a\mathcal{X}_b=	\mathcal{X}_b\mathcal{X}_a=&\mathcal{X}_{a+b}.
\end{align}
One may convince themselves by examining Eq.~\eqref{eq:xab_algebra_0} carefully.
\begin{align}\nonumber
	\label{eq:xab_algebra_0_derv}\nonumber
	\mathcal{X}_a\mathcal{X}_b=&\Big(\sum_m\ketbra{m+a}{m}\Big)\Big(\sum_n\ketbra{n+b}{n}\Big)\\\nonumber
	=&\sum_m\sum_n\ket{m+a}\braket{m}{n+b}\bra{n}\\\nonumber
	=&\sum_m\sum_n\ket{m+a}\delta_{m,n+b}\bra{n}\\\nonumber
	=&\sum_n\ketbra{(n+a)+b}{n}\\\nonumber
	=&\sum_n\ketbra{n+(a+b)}{n}\\
	=&\mathcal{X}_{a+b},
\end{align}
where $\delta_{i,j}$ is the Kronecker delta, and we also have
\begin{align}
	\label{eq:xab_algebra_1_derv}\nonumber
	\mathcal{X}_b\mathcal{X}_a=&\Big(\sum_n\ketbra{n+b}{n}\Big)\Big(\sum_m\ketbra{m+a}{m}\Big)\\\nonumber
	=&\sum_n\sum_m\ket{n+b}\braket{n}{m+a}\bra{m}\\\nonumber
	=&\sum_n\sum_m\ket{n+b}\delta_{n,m+a}\bra{m}\\\nonumber
	=&\sum_m\ketbra{(m+a)+b)}{m}\\\nonumber
	=&\sum_m\ketbra{m+(a+b)}{m}\\
	=&\mathcal{X}_{a+b}.
\end{align}

From now on, we are going to get rid of the usage of $\circ$, the symbol for denoting the composition rule of linear transformations. Therefore we can write $\hat{\mathcal{U}}_{i}^{p_j}$ in the main text as
\begin{align}
	\label{eq:u_sigma_decomp}\nonumber
	\hat{\mathcal{S}}(\hat{\mathcal{C}}_i^{p_j}\otimes\hat{\mathbb{I}}_W)=&\big(\ketbra{\vartriangleright}{\vartriangleright}\otimes\mathcal{X}_1+\ketbra{\blacktriangleleft}{\blacktriangleleft}\otimes\mathcal{X}_{-1}\big)(\hat{\mathcal{C}}_i^{p_j}\otimes\hat{\mathbb{I}}_W)\\
	=&\big(\ketbra{\vartriangleright}{\vartriangleright} \hat{\mathcal{C}}_i^{p_j}\big)\otimes\mathcal{X}_1+\big(\ketbra{\blacktriangleleft}{\blacktriangleleft} \hat{\mathcal{C}}_i^{p_j}\big)\otimes\mathcal{X}_{-1}.
\end{align}
By inserting Eq.~\eqref{eq:u_sigma_decomp} into Eq.~\eqref{eq:u_sigma01_explicit}, we have
\begin{align}
	\label{eq:usigma_expansion}\nonumber
	\hat{\mathbb{U}}^{\sigma_{j}}=&\prod_{i=\sigma_j-1}^{0}\Big[\big(\ketbra{\vartriangleright}{\vartriangleright}\hat{\mathcal{C}}_i^{p_j}\big)\otimes\mathcal{X}_1+\big(\ketbra{\blacktriangleleft}{\blacktriangleleft}\hat{\mathcal{C}}_i^{p_j}\big)\otimes\mathcal{X}_{-1}\Big]\\
	=&\sum_{k=0}^{2^{\sigma_j}-1}\mathcal{A}_k^j\otimes\mathcal{X}_{D_k},
\end{align}
where $\mathcal{A}_k^j$ is a product of $\sigma_j$ operators of $\big(\ketbra{\phi}{\phi}\hat{\mathcal{C}}_i^{p_j}\big)$, and $D_k$ is the $F_k-B_k$ if we define $F_k$ and $B_k$ as the number of appearance of $\ketbra{\vartriangleright}{\vartriangleright}$ and $\ketbra{\blacktriangleleft}{\blacktriangleleft}$ in $\mathcal{A}_k^j$ respectively. We here choose to order $\mathcal{A}_k^j$ in a natural order, such that 
\begin{subequations}
\begin{align}
	\label{eq:def_operaotr_A_a}
	\mathcal{A}_0^j=&\big(\ketbra{\vartriangleright}{\vartriangleright}\hat{\mathcal{C}}_{\sigma_j-1}^{p_j}\big)\big(\ketbra{\vartriangleright}{\vartriangleright}\hat{\mathcal{C}}_{\sigma_j-2}^{p_j}\big)\dots\big(\ketbra{\vartriangleright}{\vartriangleright}\hat{\mathcal{C}}_0^{p_j}\big)\\
	\label{eq:def_operaotr_A_b}
	\mathcal{A}_1^j=&\big(\ketbra{\vartriangleright}{\vartriangleright}\hat{\mathcal{C}}_{\sigma_j-1}^{p_j}\big)\big(\ketbra{\vartriangleright}{\vartriangleright}\hat{\mathcal{C}}_{\sigma_j-2}^{p_j}\big)\dots\big(\ketbra{\blacktriangleleft}{\blacktriangleleft}\hat{\mathcal{C}}_0^{p_j}\big)\\\nonumber
	\dots\\
	\label{eq:def_operaotr_A_c}
	\mathcal{A}_{2^{\sigma_j-1}}^j=&\big(\ketbra{\blacktriangleleft}{\blacktriangleleft}\hat{\mathcal{C}}_{\sigma_j-1}^{p_j}\big)\big(\ketbra{\blacktriangleleft}{\blacktriangleleft}\hat{\mathcal{C}}_{\sigma_j-2}^{p_j}\big)\dots\big(\ketbra{\blacktriangleleft}{\blacktriangleleft}\hat{\mathcal{C}}_0^{p_j}\big).
\end{align}
\end{subequations}
We now look back at Eq.~\eqref{eq:rho_cw_f12_derv}, and write it as
\begin{align}
	\label{eq:rho_cw_f12_final}
	\rho_{|\mathcal{F}_1^2}^{CW}=&\frac{1}{4}({\hat{\mathbb{U}}^{\sigma_1}}{\hat{\mathbb{U}}^{\sigma_0}}-{\hat{\mathbb{U}}^{\sigma_0}}{\hat{\mathbb{U}}^{\sigma_1}}){\rho}_{\star}^{CW}({\hat{\mathbb{U}}^{\sigma_0 \dagger}}{\hat{\mathbb{U}}^{\sigma_1 \dagger}}-{\hat{\mathbb{U}}^{\sigma_1 \dagger}}{\hat{\mathbb{U}}^{\sigma_0 \dagger}}).
	\end{align}
If a measurement is further made on the coin, it can be seen easily that the reduced state will be a \emph{pure} state.
\begin{align}
	\label{eq:u_sigma1minus0}\nonumber
	\hat{\mathbb{U}}^{\sigma_{1}}\hat{\mathbb{U}}^{\sigma_{0}}-\hat{\mathbb{U}}^{\sigma_{0}}\hat{\mathbb{U}}^{\sigma_{1}}=&
	\Bigg(\sum_{l=0}^{2^{\sigma_1}-1}\mathcal{A}_l^1\otimes\mathcal{X}_{D_l}\Bigg)
	\Bigg(\sum_{k=0}^{2^{\sigma_0}-1}\mathcal{A}_k^0\otimes\mathcal{X}_{D_k}\Bigg)
	-\Bigg(\sum_{k=0}^{2^{\sigma_0}-1}\mathcal{A}_k^0\otimes\mathcal{X}_{D_k}\Bigg)
	\Bigg(\sum_{l=0}^{2^{\sigma_1}-1}\mathcal{A}_l^1\otimes\mathcal{X}_{D_l}\Bigg)\\
	=&\sum_{l=0}^{{2^{\sigma_1}-1}}\sum_{k=0}^{2^{\sigma_0}-1}
\Big(\mathcal{A}_l^1\mathcal{A}_k^0-\mathcal{A}_k^0\mathcal{A}_l^1\Big)\otimes\mathcal{X}_{D_k+D_l},
\end{align}
where we have used Eq.~\eqref{eq:xab_algebra_0}, the commutativity property of the operators $\mathcal{X}_{D_k}$ and $\mathcal{X}_{D_l}$. Recall our assumption for the initial state of the coin, ${\rho}_{\star}^{C}=\ketbra{{\vartriangleright}}{{\vartriangleright}}$, and since we want to prove the symmetry property for population terms of $\rho_{|{\vartriangleright},\mathcal{F}_1^2}^W$, it suffices to show the following is true
\begin{align}
	\label{eq:theorem_1_goal}\nonumber
	\abs{\sum_{k,l}\bra{{\vartriangleright}}\mathcal{A}_l^1\mathcal{A}_k^0-\mathcal{A}_k^0\mathcal{A}_l^1\ket{{\vartriangleright}}}=&\abs{\sum_{k',l'}\bra{{\vartriangleright}}\mathcal{A}_{l'}^1\mathcal{A}_{k'}^0-\mathcal{A}_{k'}^0\mathcal{A}_{l'}^1\ket{{\vartriangleright}}},\\
	{\text{for~all}} ~k,l ,k',l'~{\text{satisfying~}}&D_k+D_l+D_{k'}+D_{l'}=0.
\end{align}

\textbf{Propisition 1:} $\sum_{k_s,l_s}\bra{{\vartriangleright}}\mathcal{A}_{l_s}^1\mathcal{A}_{k_s}^0-\mathcal{A}_{k_s}^0\mathcal{A}_{l_s}^1\ket{{\vartriangleright}}={\sum_{k_{s'},l_{s'}}\overline{\bra{{\vartriangleright}}\mathcal{A}_{l_{s'}}^1\mathcal{A}_{k_{s'
}}^0-\mathcal{A}_{k_{s'}}^0\mathcal{A}_{l_{s'}}^1\ket{{\vartriangleright}}}}$ for $\sigma_0+\sigma_1$ is even and $\sum_{k_s,l_s}\bra{{\vartriangleright}}\mathcal{A}_{l_s}^1\mathcal{A}_{k_s}^0-\mathcal{A}_{k_s}^0\mathcal{A}_{l_s}^1\ket{{\vartriangleright}}=-{\sum_{k_{s'},l_{s'}}\overline{\bra{{\vartriangleright}}\mathcal{A}_{l_{s'}}^1\mathcal{A}_{k_{s'
}}^0-\mathcal{A}_{k_{s'}}^0\mathcal{A}_{l_{s'}}^1\ket{{\vartriangleright}}}}$ for $\sigma_0+\sigma_1$ is odd, where $D_{k_s},D_{l_s}$ are chosen to be all possible combinations that add up to a same number $s$, and $s=-s'$.

\emph{Proof.---} Here we prove that Eq.~\eqref{eq:theorem_1_goal} is true by induction. For $\sigma_0+\sigma_1=1,2$, it is too trivial, thus we will choose $\sigma_0+\sigma_1=3,4$ as the base cases. 

For simplicity, let us assume that $\sigma_0=\sigma_1=2$. We write out $\hat{\mathbb{U}}^{\sigma_{1}}\hat{\mathbb{U}}^{\sigma_{0}}-\hat{\mathbb{U}}^{\sigma_{0}}\hat{\mathbb{U}}^{\sigma_{1}}$ explicitly as follows,
\begin{align}
	\label{eq:u1u0minusu0u1}\nonumber
	\hat{\mathbb{U}}^{\sigma_{1}}&\hat{\mathbb{U}}^{\sigma_{0}}-\hat{\mathbb{U}}^{\sigma_{0}}\hat{\mathbb{U}}^{\sigma_{1}}=\\\nonumber
	&~~~(\ketbra{{\vartriangleright}}{{\vartriangleright}}\hat{\mathcal{C}}_{1}^{p_1}\ketbra{{\vartriangleright}}{{\vartriangleright}}\hat{\mathcal{C}}_{0}^{p_1}\ketbra{{\vartriangleright}}{{\vartriangleright}}\hat{\mathcal{C}}_{1}^{p_0}\ketbra{{\vartriangleright}}{{\vartriangleright}}\hat{\mathcal{C}}_{0}^{p_0}-\ketbra{{\vartriangleright}}{{\vartriangleright}}\hat{\mathcal{C}}_{1}^{p_0}\ketbra{{\vartriangleright}}{{\vartriangleright}}\hat{\mathcal{C}}_{0}^{p_0}\ketbra{{\vartriangleright}}{{\vartriangleright}}\hat{\mathcal{C}}_{1}^{p_1}\ketbra{{\vartriangleright}}{{\vartriangleright}}\hat{\mathcal{C}}_{0}^{p_1})\otimes\mathcal{X}_4\\\nonumber
	&+(\ketbra{{\vartriangleright}}{{\vartriangleright}}\hat{\mathcal{C}}_{1}^{p_1}\ketbra{{\vartriangleright}}{{\vartriangleright}}\hat{\mathcal{C}}_{0}^{p_1}\ketbra{{\vartriangleright}}{{\vartriangleright}}\hat{\mathcal{C}}_{1}^{p_0}\ketbra{{\blacktriangleleft}}{{\blacktriangleleft}}\hat{\mathcal{C}}_{0}^{p_0}-\ketbra{{\vartriangleright}}{{\vartriangleright}}\hat{\mathcal{C}}_{1}^{p_0}\ketbra{{\blacktriangleleft}}{{\blacktriangleleft}}\hat{\mathcal{C}}_{0}^{p_0}\ketbra{{\vartriangleright}}{{\vartriangleright}}\hat{\mathcal{C}}_{1}^{p_1}\ketbra{{\vartriangleright}}{{\vartriangleright}}\hat{\mathcal{C}}_{0}^{p_1})\otimes\mathcal{X}_2\\\nonumber
	&+(\ketbra{{\vartriangleright}}{{\vartriangleright}}\hat{\mathcal{C}}_{1}^{p_1}\ketbra{{\vartriangleright}}{{\vartriangleright}}\hat{\mathcal{C}}_{0}^{p_1}\ketbra{{\blacktriangleleft}}{{\blacktriangleleft}}\hat{\mathcal{C}}_{1}^{p_0}\ketbra{{\vartriangleright}}{{\vartriangleright}}\hat{\mathcal{C}}_{0}^{p_0}-\ketbra{{\blacktriangleleft}}{{\blacktriangleleft}}\hat{\mathcal{C}}_{1}^{p_0}\ketbra{{\vartriangleright}}{{\vartriangleright}}\hat{\mathcal{C}}_{0}^{p_0}\ketbra{{\vartriangleright}}{{\vartriangleright}}\hat{\mathcal{C}}_{1}^{p_1}\ketbra{{\vartriangleright}}{{\vartriangleright}}\hat{\mathcal{C}}_{0}^{p_1})\otimes\mathcal{X}_2\\\nonumber
	&+(\ketbra{{\vartriangleright}}{{\vartriangleright}}\hat{\mathcal{C}}_{1}^{p_1}\ketbra{{\vartriangleright}}{{\vartriangleright}}\hat{\mathcal{C}}_{0}^{p_1}\ketbra{{\blacktriangleleft}}{{\blacktriangleleft}}\hat{\mathcal{C}}_{1}^{p_0}\ketbra{{\blacktriangleleft}}{{\blacktriangleleft}}\hat{\mathcal{C}}_{0}^{p_0}-\ketbra{{\blacktriangleleft}}{{\blacktriangleleft}}\hat{\mathcal{C}}_{1}^{p_0}\ketbra{{\blacktriangleleft}}{{\blacktriangleleft}}\hat{\mathcal{C}}_{0}^{p_0}\ketbra{{\vartriangleright}}{{\vartriangleright}}\hat{\mathcal{C}}_{1}^{p_1}\ketbra{{\vartriangleright}}{{\vartriangleright}}\hat{\mathcal{C}}_{0}^{p_1})\otimes\mathcal{X}_0\\\nonumber
	&+(\ketbra{{\vartriangleright}}{{\vartriangleright}}\hat{\mathcal{C}}_{1}^{p_1}\ketbra{{\blacktriangleleft}}{{\blacktriangleleft}}\hat{\mathcal{C}}_{0}^{p_1}\ketbra{{\vartriangleright}}{{\vartriangleright}}\hat{\mathcal{C}}_{1}^{p_0}\ketbra{{\vartriangleright}}{{\vartriangleright}}\hat{\mathcal{C}}_{0}^{p_0}-\ketbra{{\vartriangleright}}{{\vartriangleright}}\hat{\mathcal{C}}_{1}^{p_0}\ketbra{{\vartriangleright}}{{\vartriangleright}}\hat{\mathcal{C}}_{0}^{p_0}\ketbra{{\vartriangleright}}{{\vartriangleright}}\hat{\mathcal{C}}_{1}^{p_1}\ketbra{{\blacktriangleleft}}{{\blacktriangleleft}}\hat{\mathcal{C}}_{0}^{p_1})\otimes\mathcal{X}_2\\\nonumber
	&+(\ketbra{{\vartriangleright}}{{\vartriangleright}}\hat{\mathcal{C}}_{1}^{p_1}\ketbra{{\blacktriangleleft}}{{\blacktriangleleft}}\hat{\mathcal{C}}_{0}^{p_1}\ketbra{{\vartriangleright}}{{\vartriangleright}}\hat{\mathcal{C}}_{1}^{p_0}\ketbra{{\blacktriangleleft}}{{\blacktriangleleft}}\hat{\mathcal{C}}_{0}^{p_0}-\ketbra{{\vartriangleright}}{{\vartriangleright}}\hat{\mathcal{C}}_{1}^{p_0}\ketbra{{\blacktriangleleft}}{{\blacktriangleleft}}\hat{\mathcal{C}}_{0}^{p_0}\ketbra{{\vartriangleright}}{{\vartriangleright}}\hat{\mathcal{C}}_{1}^{p_1}\ketbra{{\blacktriangleleft}}{{\blacktriangleleft}}\hat{\mathcal{C}}_{0}^{p_1})\otimes\mathcal{X}_0\\\nonumber
	&+(\ketbra{{\vartriangleright}}{{\vartriangleright}}\hat{\mathcal{C}}_{1}^{p_1}\ketbra{{\blacktriangleleft}}{{\blacktriangleleft}}\hat{\mathcal{C}}_{0}^{p_1}\ketbra{{\blacktriangleleft}}{{\blacktriangleleft}}\hat{\mathcal{C}}_{1}^{p_0}\ketbra{{\vartriangleright}}{{\vartriangleright}}\hat{\mathcal{C}}_{0}^{p_0}-\ketbra{{\blacktriangleleft}}{{\blacktriangleleft}}\hat{\mathcal{C}}_{1}^{p_0}\ketbra{{\vartriangleright}}{{\vartriangleright}}\hat{\mathcal{C}}_{0}^{p_0}\ketbra{{\vartriangleright}}{{\vartriangleright}}\hat{\mathcal{C}}_{1}^{p_1}\ketbra{{\blacktriangleleft}}{{\blacktriangleleft}}\hat{\mathcal{C}}_{0}^{p_1})\otimes\mathcal{X}_0\\\nonumber
	&+(\ketbra{{\vartriangleright}}{{\vartriangleright}}\hat{\mathcal{C}}_{1}^{p_1}\ketbra{{\blacktriangleleft}}{{\blacktriangleleft}}\hat{\mathcal{C}}_{0}^{p_1}\ketbra{{\blacktriangleleft}}{{\blacktriangleleft}}\hat{\mathcal{C}}_{1}^{p_0}\ketbra{{\blacktriangleleft}}{{\blacktriangleleft}}\hat{\mathcal{C}}_{0}^{p_0}-\ketbra{{\blacktriangleleft}}{{\blacktriangleleft}}\hat{\mathcal{C}}_{1}^{p_0}\ketbra{{\blacktriangleleft}}{{\blacktriangleleft}}\hat{\mathcal{C}}_{0}^{p_0}\ketbra{{\vartriangleright}}{{\vartriangleright}}\hat{\mathcal{C}}_{1}^{p_1}\ketbra{{\blacktriangleleft}}{{\blacktriangleleft}}\hat{\mathcal{C}}_{0}^{p_1})\otimes\mathcal{X}_{-2}\\\nonumber
	&+(\ketbra{{\blacktriangleleft}}{{\blacktriangleleft}}\hat{\mathcal{C}}_{1}^{p_1}\ketbra{{\vartriangleright}}{{\vartriangleright}}\hat{\mathcal{C}}_{0}^{p_1}\ketbra{{\vartriangleright}}{{\vartriangleright}}\hat{\mathcal{C}}_{1}^{p_0}\ketbra{{\vartriangleright}}{{\vartriangleright}}\hat{\mathcal{C}}_{0}^{p_0}-\ketbra{{\vartriangleright}}{{\vartriangleright}}\hat{\mathcal{C}}_{1}^{p_0}\ketbra{{\vartriangleright}}{{\vartriangleright}}\hat{\mathcal{C}}_{0}^{p_0}\ketbra{{\blacktriangleleft}}{{\blacktriangleleft}}\hat{\mathcal{C}}_{1}^{p_1}\ketbra{{\vartriangleright}}{{\vartriangleright}}\hat{\mathcal{C}}_{0}^{p_1})\otimes\mathcal{X}_2\\\nonumber
	&+(\ketbra{{\blacktriangleleft}}{{\blacktriangleleft}}\hat{\mathcal{C}}_{1}^{p_1}\ketbra{{\vartriangleright}}{{\vartriangleright}}\hat{\mathcal{C}}_{0}^{p_1}\ketbra{{\vartriangleright}}{{\vartriangleright}}\hat{\mathcal{C}}_{1}^{p_0}\ketbra{{\blacktriangleleft}}{{\blacktriangleleft}}\hat{\mathcal{C}}_{0}^{p_0}-\ketbra{{\vartriangleright}}{{\vartriangleright}}\hat{\mathcal{C}}_{1}^{p_0}\ketbra{{\blacktriangleleft}}{{\blacktriangleleft}}\hat{\mathcal{C}}_{0}^{p_0}\ketbra{{\blacktriangleleft}}{{\blacktriangleleft}}\hat{\mathcal{C}}_{1}^{p_1}\ketbra{{\vartriangleright}}{{\vartriangleright}}\hat{\mathcal{C}}_{0}^{p_1})\otimes\mathcal{X}_0\\\nonumber
	&+(\ketbra{{\blacktriangleleft}}{{\blacktriangleleft}}\hat{\mathcal{C}}_{1}^{p_1}\ketbra{{\vartriangleright}}{{\vartriangleright}}\hat{\mathcal{C}}_{0}^{p_1}\ketbra{{\blacktriangleleft}}{{\blacktriangleleft}}\hat{\mathcal{C}}_{1}^{p_0}\ketbra{{\vartriangleright}}{{\vartriangleright}}\hat{\mathcal{C}}_{0}^{p_0}-\ketbra{{\blacktriangleleft}}{{\blacktriangleleft}}\hat{\mathcal{C}}_{1}^{p_0}\ketbra{{\vartriangleright}}{{\vartriangleright}}\hat{\mathcal{C}}_{0}^{p_0}\ketbra{{\blacktriangleleft}}{{\blacktriangleleft}}\hat{\mathcal{C}}_{1}^{p_1}\ketbra{{\vartriangleright}}{{\vartriangleright}}\hat{\mathcal{C}}_{0}^{p_1})\otimes\mathcal{X}_0\\\nonumber
	&+(\ketbra{{\blacktriangleleft}}{{\blacktriangleleft}}\hat{\mathcal{C}}_{1}^{p_1}\ketbra{{\vartriangleright}}{{\vartriangleright}}\hat{\mathcal{C}}_{0}^{p_1}\ketbra{{\blacktriangleleft}}{{\blacktriangleleft}}\hat{\mathcal{C}}_{1}^{p_0}\ketbra{{\blacktriangleleft}}{{\blacktriangleleft}}\hat{\mathcal{C}}_{0}^{p_0}-\ketbra{{\blacktriangleleft}}{{\blacktriangleleft}}\hat{\mathcal{C}}_{1}^{p_0}\ketbra{{\blacktriangleleft}}{{\blacktriangleleft}}\hat{\mathcal{C}}_{0}^{p_0}\ketbra{{\blacktriangleleft}}{{\blacktriangleleft}}\hat{\mathcal{C}}_{1}^{p_1}\ketbra{{\vartriangleright}}{{\vartriangleright}}\hat{\mathcal{C}}_{0}^{p_1})\otimes\mathcal{X}_{-2}\\\nonumber
	&+(\ketbra{{\blacktriangleleft}}{{\blacktriangleleft}}\hat{\mathcal{C}}_{1}^{p_1}\ketbra{{\blacktriangleleft}}{{\blacktriangleleft}}\hat{\mathcal{C}}_{0}^{p_1}\ketbra{{\vartriangleright}}{{\vartriangleright}}\hat{\mathcal{C}}_{1}^{p_0}\ketbra{{\vartriangleright}}{{\vartriangleright}}\hat{\mathcal{C}}_{0}^{p_0}-\ketbra{{\vartriangleright}}{{\vartriangleright}}\hat{\mathcal{C}}_{1}^{p_0}\ketbra{{\vartriangleright}}{{\vartriangleright}}\hat{\mathcal{C}}_{0}^{p_0}\ketbra{{\blacktriangleleft}}{{\blacktriangleleft}}\hat{\mathcal{C}}_{1}^{p_1}\ketbra{{\blacktriangleleft}}{{\blacktriangleleft}}\hat{\mathcal{C}}_{0}^{p_1})\otimes\mathcal{X}_0\\\nonumber
	&+(\ketbra{{\vartriangleright}}{{\vartriangleright}}\hat{\mathcal{C}}_{1}^{p_1}\ketbra{{\blacktriangleleft}}{{\blacktriangleleft}}\hat{\mathcal{C}}_{0}^{p_1}\ketbra{{\vartriangleright}}{{\vartriangleright}}\hat{\mathcal{C}}_{1}^{p_0}\ketbra{{\blacktriangleleft}}{{\blacktriangleleft}}\hat{\mathcal{C}}_{0}^{p_0}-\ketbra{{\vartriangleright}}{{\vartriangleright}}\hat{\mathcal{C}}_{1}^{p_0}\ketbra{{\blacktriangleleft}}{{\blacktriangleleft}}\hat{\mathcal{C}}_{0}^{p_0}\ketbra{{\vartriangleright}}{{\vartriangleright}}\hat{\mathcal{C}}_{1}^{p_1}\ketbra{{\blacktriangleleft}}{{\blacktriangleleft}}\hat{\mathcal{C}}_{0}^{p_1})\otimes\mathcal{X}_{-2}\\\nonumber
	&+(\ketbra{{\blacktriangleleft}}{{\blacktriangleleft}}\hat{\mathcal{C}}_{1}^{p_1}\ketbra{{\blacktriangleleft}}{{\blacktriangleleft}}\hat{\mathcal{C}}_{0}^{p_1}\ketbra{{\blacktriangleleft}}{{\blacktriangleleft}}\hat{\mathcal{C}}_{1}^{p_0}\ketbra{{\vartriangleright}}{{\vartriangleright}}\hat{\mathcal{C}}_{0}^{p_0}-\ketbra{{\blacktriangleleft}}{{\blacktriangleleft}}\hat{\mathcal{C}}_{1}^{p_0}\ketbra{{\vartriangleright}}{{\vartriangleright}}\hat{\mathcal{C}}_{0}^{p_0}\ketbra{{\blacktriangleleft}}{{\blacktriangleleft}}\hat{\mathcal{C}}_{1}^{p_1}\ketbra{{\blacktriangleleft}}{{\blacktriangleleft}}\hat{\mathcal{C}}_{0}^{p_1})\otimes\mathcal{X}_{-2}\\
	&+(\ketbra{{\blacktriangleleft}}{{\blacktriangleleft}}\hat{\mathcal{C}}_{1}^{p_1}\ketbra{{\blacktriangleleft}}{{\blacktriangleleft}}\hat{\mathcal{C}}_{0}^{p_1}\ketbra{{\blacktriangleleft}}{{\blacktriangleleft}}\hat{\mathcal{C}}_{1}^{p_0}\ketbra{{\blacktriangleleft}}{{\blacktriangleleft}}\hat{\mathcal{C}}_{0}^{p_0}-\ketbra{{\blacktriangleleft}}{{\blacktriangleleft}}\hat{\mathcal{C}}_{1}^{p_0}\ketbra{{\blacktriangleleft}}{{\blacktriangleleft}}\hat{\mathcal{C}}_{0}^{p_0}\ketbra{{\blacktriangleleft}}{{\blacktriangleleft}}\hat{\mathcal{C}}_{1}^{p_1}\ketbra{{\blacktriangleleft}}{{\blacktriangleleft}}\hat{\mathcal{C}}_{0}^{p_1})\otimes\mathcal{X}_{-4}.
\end{align}
Contracting Eq.~\eqref{eq:u1u0minusu0u1} using $\bra{\vartriangleright}\dots\ket{\vartriangleright}$, we find that the amplitudes for $\mathcal{X}_2$ and $\mathcal{X}_{-2}$ are complex conjugate with respect to each other, and for $\mathcal{X}_4$ and $\mathcal{X}_{-4}$, the contribution are zero. One can also confirm it is true also for the case of $\sigma_0+\sigma_1=3$. 
Next consider $\sigma_0+\sigma_1+1$, where $\sigma_0+\sigma_1$ is odd. 

Next, we want to compute the amplitude for the walker at the position $D_{k_t}+D_{l_t}$ and $D_{k_{t'}}+D_{l_{t'}}$, where $t=-t'$. Let us give them the names $P_t$ and $P_t'$. These two probabilities  can be computed in the following way
\begin{align}
	\label{eq:dkdl_prob_0}\nonumber
	P_t=&\\
	\sum_{k_{t-1},l_{t-1}}\bra{{\vartriangleright}}\mathcal{A}_{l_{t-1}}^1\mathcal{A}_{k_{t-1}}^0-\mathcal{A}_{k_{t-1}}^0\mathcal{A}_{l_{t-1}}^1\ket{{\vartriangleright}}\mathbf{Z}~~+&\sum_{k_{t+1},l_{t+1}}\bra{{\vartriangleright}}\mathcal{A}_{l_{t+1}}^1\mathcal{A}_{k_{t+1}}^0-\mathcal{A}_{k_{t+1}}^0\mathcal{A}_{l_{t+1}}^1\ket{{\vartriangleright}}(-\overline{\mathbf{Z}})\\
	=&\mathbf{X}\cdot\mathbf{Z}+\mathbf{Y}\cdot(-\overline{\mathbf{Z}})
\end{align}
and
\begin{align}
	\label{eq:dkdl_prob_1}\nonumber
	P_t'=&\\
	\sum_{k_{t'-1},l_{t'-1}}\bra{{\vartriangleright}}\mathcal{A}_{l_{t'-1}}^1\mathcal{A}_{k_{t'-1}}^0-\mathcal{A}_{k_{t'-1}}^0\mathcal{A}_{l_{t'-1}}^1\ket{{\vartriangleright}}\mathbf{Z}~~+&\sum_{k_{t'+1},l_{t'+1}}\bra{{\vartriangleright}}\mathcal{A}_{l_{t'+1}}^1\mathcal{A}_{k_{t'+1}}^0-\mathcal{A}_{k_{t'+1}}^0\mathcal{A}_{l_{t'+1}}^1\ket{{\vartriangleright}}(-\overline{\mathbf{Z}})\\
	=&(-\overline{\mathbf{Y}})\cdot\mathbf{Z}+(-\overline{\mathbf{X}})\cdot(-\overline{\mathbf{Z}}).
\end{align}
In the above, we have used the induction hypothesis and the reason for introducing the complex number $\mathbf{Z}$ can be understood by recalling our definition of the SU(2) coin operators. Clearly, Eq.~\eqref{eq:dkdl_prob_0} is exactly the conjugate of Eq.~\eqref{eq:dkdl_prob_1}. 

By the similar procedure, that is start from the assumption that where $\sigma_0+\sigma_1$ is even, one find that the resulting two new numbers are exactly complex conjugate with respect to each other. Because the sign before a complex number is irrelevant to the probability amplitude, we make the conclusion that Eq.~\eqref{eq:theorem_1_goal} is true, which then implies that Theorem 1 is true.
\section{Proof of Lemma 1}
The proof of Lemma 1 is straightforward. Let us add up Eq.~\eqref{eq:rho_cw_f02_derv} and Eq.~\eqref{eq:rho_cw_f12_derv}, so we have
\begin{align}
	\label{eq:lemma_1_proof}\nonumber
	\rho_{|\mathcal{F}_0^2}^{CW}+\rho_{|\mathcal{F}_1^2}^{CW}=&
	\frac{1}{4}({\hat{\mathbb{U}}^{\sigma_1}}{\hat{\mathbb{U}}^{\sigma_0}}{\rho}_{\star}^{CW}{\hat{\mathbb{U}}^{\sigma_0 \dagger}}{\hat{\mathbb{U}}^{\sigma_1 \dagger}}+{\hat{\mathbb{U}}^{\sigma_1}}{\hat{\mathbb{U}}^{\sigma_0}}{\rho}_{\star}^{CW}{\hat{\mathbb{U}}^{\sigma_1 \dagger}}{\hat{\mathbb{U}}^{\sigma_0 \dagger}}\\\nonumber
	&~{+\hat{\mathbb{U}}^{\sigma_0}}{\hat{\mathbb{U}}^{\sigma_1}}{\rho}_{\star}^{CW}{\hat{\mathbb{U}}^{\sigma_0 \dagger}}{\hat{\mathbb{U}}^{\sigma_1 \dagger}}+{\hat{\mathbb{U}}^{\sigma_0}}{\hat{\mathbb{U}}^{\sigma_1}}{\rho}_{\star}^{CW}{\hat{\mathbb{U}}^{\sigma_1 \dagger}}{\hat{\mathbb{U}}^{\sigma_0 \dagger}})\\\nonumber
	+&\frac{1}{4}({\hat{\mathbb{U}}^{\sigma_1}}{\hat{\mathbb{U}}^{\sigma_0}}{\rho}_{\star}^{CW}{\hat{\mathbb{U}}^{\sigma_0 \dagger}}{\hat{\mathbb{U}}^{\sigma_1 \dagger}}-{\hat{\mathbb{U}}^{\sigma_1}}{\hat{\mathbb{U}}^{\sigma_0}}{\rho}_{\star}^{CW}{\hat{\mathbb{U}}^{\sigma_1 \dagger}}{\hat{\mathbb{U}}^{\sigma_0 \dagger}}\\\nonumber
	&~{-\hat{\mathbb{U}}^{\sigma_0}}{\hat{\mathbb{U}}^{\sigma_1}}{\rho}_{\star}^{CW}{\hat{\mathbb{U}}^{\sigma_0 \dagger}}{\hat{\mathbb{U}}^{\sigma_1 \dagger}}+{\hat{\mathbb{U}}^{\sigma_0}}{\hat{\mathbb{U}}^{\sigma_1}}{\rho}_{\star}^{CW}{\hat{\mathbb{U}}^{\sigma_1 \dagger}}{\hat{\mathbb{U}}^{\sigma_0 \dagger}})\\
	=&\frac{1}{2}({\hat{\mathbb{U}}^{\sigma_1}}{\hat{\mathbb{U}}^{\sigma_0}}{\rho}_{\star}^{CW}{\hat{\mathbb{U}}^{\sigma_0 \dagger}}{\hat{\mathbb{U}}^{\sigma_1 \dagger}}+{\hat{\mathbb{U}}^{\sigma_0}}{\hat{\mathbb{U}}^{\sigma_1}}{\rho}_{\star}^{CW}{\hat{\mathbb{U}}^{\sigma_1 \dagger}}{\hat{\mathbb{U}}^{\sigma_0 \dagger}}),
\end{align}
which is exactly what we expected.
\section{Proof of Corollary 1}
As before, without loss of generality we assume that the initial state of the walker is $\ketbra{0}{0}$ and the coin is initialized to be in the state $\ketbra{{\vartriangleright}}{{\vartriangleright}}$. Also, for simplicity we first consider the condition of $\sigma_0=\sigma_1$.

The proof is by considering Theorem 1 and Lemma 1. By Lemma 1 we have
\begin{align}
	\label{eq:corollary_1_proof}
	\bra{n_0-l}\varrho_{0|{\vartriangleright}}^W+\varrho_{1|{\vartriangleright}}^W\ket{n_0-l}=\bra{n_0-l}\rho_{|{\vartriangleright},\mathcal{F}_0^2}^W+\rho_{|{\vartriangleright},\mathcal{F}_{1}^2}^W\ket{n_0-l}=0~~~
	\text{for~}\forall~l>0.
\end{align}
Thus $\bra{n_0-l}\rho_{|{\vartriangleright},\mathcal{F}_0^2}^W\ket{n_0-l}=-\bra{n_0-l}\rho_{|{\vartriangleright},\mathcal{F}_{1}^2}^W\ket{n_0-l}$. Also, based on the fact that for any density operator, its population terms always stay non-negative, which means that $\bra{n_0-l}\rho_{|{\vartriangleright},\mathcal{F}_{1}^2}^W\ket{n_0-l}=0$. However Theorem 1 states that the population terms in $\rho_{|{\vartriangleright},\mathcal{F}_1^2}^W$ must be symmetric with respect to $\bra{n_0}\rho_{|{\vartriangleright},\mathcal{F}_1^2}^W\ket{n_0}$, i.e., $\bra{n_0+l}\rho_{|{\vartriangleright},\mathcal{F}_{1}^2}^W\ket{n_0+l}=0$, for $\forall~l >0$.

Therefore what remains be done is to show that $\bra{0}\rho_{|{\vartriangleright},\mathcal{F}_1^2}^W\ket{0}=0$. For cases where $\sigma_0+\sigma_1$ is an odd number it is trivial since $\bra{0}\varrho_{0|{\vartriangleright}}^W+\varrho_{1|{\vartriangleright}}^W\ket{0}=0$. Whereas for even--step walks, since the probability amplitude of finding a walker at $\ket{0}$ is contributed by that of $\ket{1}$ and $\ket{-1}$ in the previous (even) number of steps, which are both zero. Finally, when $\sigma_1>\sigma_0$, because the state of non-zero probability are simply propagated farther in the forward direction, the same analysis still applies. Hence we completed the proof.
\section{Proof of Lemma 2}
\label{section:lemma_2}
To begin the proof of Lemma 2, consider two pure states, ${\hat{\mathbb{U}}^{\sigma_1}}{\hat{\mathbb{U}}^{\sigma_0}}\ket{{\vartriangleright}}\otimes\ket{0}$ and ${\hat{\mathbb{U}}^{\sigma_0}}{\hat{\mathbb{U}}^{\sigma_1}}\ket{{\vartriangleright}}\otimes\ket{0}$. Let us give them names $\ket{\Psi_0}$ and $\ket{\Psi_1}$. After a projective measurement made on the coin, $\ket{\Psi_0}$ may collapse to $\ket{\psi_0}=\bra{{\blacktriangleleft}}\ket{\Psi_0}$, which is of our interest. Then $\bra{i}\varrho_{0|\phi}^W\ket{i}=\abs{\bra{i}\ket{\psi_0}}^2$, so as for $\ket{\psi_1}=\bra{{\blacktriangleleft}}\ket{\Psi_1}$. We claim that there is no overlap shared by $\ket{\psi_0}$ and $\ket{\psi_1}$.

The key step is to show the above is true, recall that $\hat{\mathcal{D}}_i^{p_1}=\hat{\mathcal{C}}(\alpha_i^{p_1},\beta_i^{p_1},0)$, which means that any term of $\mathcal{A}_l^1\mathcal{A}_k^0$ which begins with $\ketbra{{\vartriangleright}}{{\vartriangleright}}$ will not contribute to the probability amplitude of $\ket{\psi_0}$. So does operators $\mathcal{A}_l^1\mathcal{A}_k^0$ of the form containing terms like $\ketbra{{\vartriangleright}}{{\vartriangleright}}\hat{\mathcal{D}}_i^{p_1}\ketbra{{\blacktriangleleft}}{{\blacktriangleleft}}$ due to the fact that $\bra{{\vartriangleright}}\hat{\mathcal{D}}_i^{p_1}\ket{{\blacktriangleleft}}=\bra{{\blacktriangleleft}}\hat{\mathcal{D}}_i^{p_1}\ket{{\vartriangleright}}=0$. However, only terms such as 
\begin{align}
	\label{eq:lemma_2_ex1}
	\ketbra{{\blacktriangleleft}}{{\blacktriangleleft}}\hat{\mathcal{D}}_{\sigma_1-1}^{p_1}\dots\ketbra{{\blacktriangleleft}}{{\blacktriangleleft}}\hat{\mathcal{D}}_{0}^{p_1}\dots\ketbra{{\blacktriangleleft}}{{\blacktriangleleft}}\hat{\mathcal{C}}_{\sigma_0-1}^{p_0}\dots\hat{\mathcal{C}}_{0}^{p_0}
\end{align}
have non-zero contributions, and there are $2^{\sigma_0-1}$ such terms.

For $\ket{\psi_1}$, by the same reason we are led to conclude that only $2^{\sigma_0-1}$ terms of $\mathcal{A}_k^0\mathcal{A}_l^1$ are relevant. These terms are of the form like
\begin{align}
	\label{eq:lemma_2_ex2}
\ketbra{{\blacktriangleleft}}{{\blacktriangleleft}}\hat{\mathcal{C}}_{\sigma_0-1}^{p_0}\dots\hat{\mathcal{C}}_{0}^{p_0}\ketbra{{\vartriangleright}}{{\vartriangleright}}\hat{\mathcal{D}}_{\sigma_1-1}^{p_1}\ketbra{{\vartriangleright}}{{\vartriangleright}}\dots\ketbra{{\vartriangleright}}{{\vartriangleright}}\hat{\mathcal{D}}_{0}^{p_1}.
\end{align}

We now consider the $\{\mathcal{X}_m^j\}$ operators associated with Eq.~\eqref{eq:lemma_2_ex1} and Eq.~\eqref{eq:lemma_2_ex2}. The point here is to examine the smallest index $m$ of $\{\mathcal{X}_m^0\}$ for Eq.~\eqref{eq:lemma_2_ex2} and the biggest one in $\{\mathcal{X}_m^1\}$ for Eq.~\eqref{eq:lemma_2_ex1}. It the turns out these two index are $\sigma_1-\sigma_0-2$ and $\sigma_0-\sigma_1-2$ respectively. Hence $\ket{\psi_0}$ and $\ket{\psi_1}$ have no common components whenever the following condition is satisfied,
\begin{align}
	\label{eq:proof_lemma2_inequality}\nonumber
	\sigma_1-\sigma_0-2&\geq\sigma_0-\sigma_1-2\\
	\Rightarrow ~~~~~~~~\sigma_1&\geq\sigma_0
\end{align}
 Because as long as Eq.~\eqref{eq:proof_lemma2_inequality} holds, Eq.~\eqref{eq:lemma_2_ex1} and Eq.~\eqref{eq:lemma_2_ex2} are associated with different $\mathcal{X}_i\ket{0}$. To be more specific, if we write $\ket{\psi_0}=\sum_{i}c_i\mathcal{X}_i^0\ket{0}$ and $\ket{\psi_1}=\sum_{j}d_j\mathcal{X}_j^1\ket{0}$, then $\mathcal{X}_i^0\neq\mathcal{X}_j^1$, for $\forall~i,j$.
 
Finally, to conclude we have $\rho_{|\phi,\mathcal{F}_0^2}^W$ being the density operator for $\ket{\psi_0}+\ket{\psi_1}$ and the same for $\rho_{|\phi,\mathcal{F}_1^2}^W$ and $\ket{\psi_0}-\ket{\psi_1}$ up to a common normalization factor, which then completes our proof.
\section{Proof of Corollary 2}
Since we have already shown the proof of both Lemma 1 and Lemma 2. One substitutes $\rho_{|\phi,\mathcal{F}_0^2}^W$ or $\rho_{|\phi,\mathcal{F}_1^2}^W$ into Eq.~(8) and arrives at
\begin{align}
	\label{eq:proof_corollary_2}\nonumber
	\rho_{|\phi,\mathcal{F}_0^2}^W + \rho_{|\phi,\mathcal{F}_0^2}^W& = \frac{1}{2}(\varrho_{0|\phi}^W+\varrho_{1|\phi}^W)\\
	\Rightarrow ~~~~~\rho_{|\phi,\mathcal{F}_0^2}^W =& \frac{1}{4}(\varrho_{0|\phi}^W+\varrho_{1|\phi}^W)
\end{align}
obviously it is the same for $\rho_{|\phi,\mathcal{F}_1^2}^W$.
\section{Proof of Lemma 3}
The proof of Lemma 3 is by considering $\ket{\psi_0}$ and $\ket{\psi_1}$ in Section~\ref{section:lemma_2}. Because their components in the position basis are related by the contraction of terms like Eq.~\eqref{eq:lemma_2_ex1} and Eq.~\eqref{eq:lemma_2_ex2}, and recalling the definition of $\hat{\mathcal{D}}_{i}^{p_1}$, one notices that only the pattern of operators $\ketbra{{\phi}}{{\phi}}$ in front of $\hat{\mathcal{C}}_{i}^{p_0}$ are relevant, which vary in the same way for both $\ket{\psi_0}$ and $\ket{\psi_1}$. In addition, since for each $\mathcal{X}_m$ with different $m$ there is only one term $\mathcal{A}_k^j\mathcal{A}_l^{1-j}$ contributes, one concludes that the non-zero amplitude components are identical up to a spatial translation. To obtain the length of shift, we consider the two corresponding operators that account for amplitudes of backward-most positions for $\ket{\psi_0}$ and $\ket{\psi_1}$. It is easy to see that they are
\begin{subequations}
\begin{align}
	\ketbra{{\blacktriangleleft}}{{\blacktriangleleft}}\hat{\mathcal{C}}_{\sigma_0-1}^{p_0}\dots\ketbra{{\blacktriangleleft}}{{\blacktriangleleft}}\hat{\mathcal{C}}_{0}^{p_0}\ketbra{{\vartriangleright}}{{\vartriangleright}}\hat{\mathcal{D}}_{\sigma_1-1}^{p_1}\dots\ketbra{{\vartriangleright}}{{\vartriangleright}}\hat{\mathcal{D}}_{0}^{p_1},\\
	\ketbra{{\blacktriangleleft}}{{\blacktriangleleft}}\hat{\mathcal{D}}_{\sigma_1-1}^{p_1}\dots\ketbra{{\blacktriangleleft}}{{\blacktriangleleft}}\hat{\mathcal{D}}_{0}^{p_1}\ketbra{{\blacktriangleleft}}{{\blacktriangleleft}}\hat{\mathcal{C}}_{\sigma_0-1}^{p_0}\dots\ketbra{{\blacktriangleleft}}{{\blacktriangleleft}}\hat{\mathcal{C}}_{0}^{p_0}.
\end{align}
\end{subequations}
In the above, one finds that the spatial difference between them is
\begin{align}
	-\sigma_0+\sigma_1-(-\sigma_1-\sigma_0)=2\sigma_1.
\end{align}
Hence we have $\bra{x}\varrho_{1|\phi^{\bot}_0}^W\ket{x}=\bra{x-2\sigma_1}\varrho_{0|\phi^{\bot}_0}^W\ket{x-2\sigma_1}$.
\section{Proof of Theorem 2}
We first write the initial state of the total system as follows,
\begin{align}
	\label{eq:rho_ocw_derv_N}\nonumber
	{\rho}^{\mathcal{O}CW}=&\hat{\mathbb{W}}_{\mathcal{N}}^{}{\rho}_{\star}^{\mathcal{O}CW}\hat{\mathbb{W}}_{\mathcal{N}}^{}\\\nonumber
	=&\sum_{n=0}^{\mathcal{N}-1}\ketbra{n}{n}^\mathcal{O}\otimes\mathbb{L}_n\big\{{\hat{\mathbb{U}}^{\sigma_{\mathcal{N}-1}}}\dots{\hat{\mathbb{U}}^{\sigma_0}}\big\}\Big({\rho}_{\star}^{\mathcal{O}}\otimes{\rho}_{\star}^{CW}\Big)\sum_{n'=0}^{\mathcal{N}-1}\ketbra{n'}{n'}^\mathcal{O}\otimes\mathbb{L}_{n'}\big\{{\hat{\mathbb{U}}^{\sigma_{\mathcal{N}-1}}}\dots{\hat{\mathbb{U}}^{\sigma_0}}\big\}^{\dagger}\\\nonumber
	=&\sum_{n=0}^{\mathcal{N}-1}\sum_{n'=0}^{\mathcal{N}-1}\Big(\ketbra{n}{n}^\mathcal{O} \rho_{\star}^{\mathcal{O}}\ketbra{n'}{n'}^\mathcal{O}\Big)\otimes \mathbb{L}_n\big\{{\hat{\mathbb{U}}^{\sigma_{\mathcal{N}-1}}}\dots{\hat{\mathbb{U}}^{\sigma_0}}\big\}{\rho}_{\star}^{CW}\mathbb{L}_{n'}\big\{{\hat{\mathbb{U}}^{\sigma_{\mathcal{N}-1}}}\dots{\hat{\mathbb{U}}^{\sigma_0}}\big\}^{\dagger}\\\nonumber
	=&\frac{1}{\mathcal{N}}\sum_{j=0}^{\mathcal{N}-1}\sum_{j'=0}^{\mathcal{N}-1}\sum_{n=0}^{\mathcal{N}-1}\sum_{n'=0}^{\mathcal{N}-1}\Big(\ketbra{n}{n}^\mathcal{O} \ketbra{j}{j'}^{\mathcal{O}}\ketbra{n'}{n'}^\mathcal{O}\Big)\otimes \mathbb{L}_n\big\{{\hat{\mathbb{U}}^{\sigma_{\mathcal{N}-1}}}\dots{\hat{\mathbb{U}}^{\sigma_0}}\big\}{\rho}_{\star}^{CW}\mathbb{L}_{n'}\big\{{\hat{\mathbb{U}}^{\sigma_{\mathcal{N}-1}}}\dots{\hat{\mathbb{U}}^{\sigma_0}}\big\}^{\dagger}\\\nonumber
	=&\frac{1}{\mathcal{N}}\sum_{j=0}^{\mathcal{N}-1}\sum_{j'=0}^{\mathcal{N}-1}\sum_{n=0}^{\mathcal{N}-1}\sum_{n'=0}^{\mathcal{N}-1}\Big(\ketbra{n}{n'}^\mathcal{O}\delta_{n,j}\delta_{j',n'}\Big)\otimes \mathbb{L}_n\big\{{\hat{\mathbb{U}}^{\sigma_{\mathcal{N}-1}}}\dots{\hat{\mathbb{U}}^{\sigma_0}}\big\}{\rho}_{\star}^{CW}\mathbb{L}_{n'}\big\{{\hat{\mathbb{U}}^{\sigma_{\mathcal{N}-1}}}\dots{\hat{\mathbb{U}}^{\sigma_0}}\big\}^{\dagger}\\\nonumber
	=&\frac{1}{\mathcal{N}}\sum_{n=0}^{\mathcal{N}-1}\sum_{n'=0}^{\mathcal{N}-1}\Big(\ketbra{n}{n'}^\mathcal{O}\Big)\otimes \mathbb{L}_n\big\{{\hat{\mathbb{U}}^{\sigma_{\mathcal{N}-1}}}\dots{\hat{\mathbb{U}}^{\sigma_0}}\big\}{\rho}_{\star}^{CW}\mathbb{L}_{n'}\big\{{\hat{\mathbb{U}}^{\sigma_{\mathcal{N}-1}}}\dots{\hat{\mathbb{U}}^{\sigma_0}}\big\}^{\dagger}\\
	=&\frac{1}{\mathcal{N}}\sum_{n=0}^{\mathcal{N}-1}\sum_{n'=0}^{\mathcal{N}-1}\Big(\ketbra{n}{n'}^\mathcal{O}\Big)\otimes \mathbb{L}_n\big\{\ast\big\}{\rho}_{\star}^{CW}\mathbb{L}_{n'}\big\{\ast\big\}^{\dagger}.
\end{align}
where in the final step we use the shorthand $\mathbb{L}_n\big\{\ast\big\}$ to represent $\mathbb{L}_n\big\{{\hat{\mathbb{U}}^{\sigma_{\mathcal{N}-1}}}\dots{\hat{\mathbb{U}}^{\sigma_0}}\big\}$. Now we use Eq.~\eqref{eq:rho_ocw_derv_N} to obtain $\rho_{|\mathcal{F}_m^\mathcal{N}}^{CW}$,
\begin{align}
	\label{eq:rho_cw_fmN_derv}\nonumber
	\rho_{|\mathcal{F}_m^\mathcal{N}}^{CW}=&\bra{\mathcal{F}_m^\mathcal{N}}\rho^{\mathcal{O}CW}\ket{\mathcal{F}_m^\mathcal{N}}\\\nonumber
	=&\frac{1}{\sqrt{\mathcal{N}}}\sum_ke^{i\frac{2\pi m}{\mathcal{N}}k}\bra{k}^{\mathcal{O}}\rho^{\mathcal{O}CW}\frac{1}{\sqrt{\mathcal{N}}}\sum_{k'}e^{-i\frac{2\pi m}{\mathcal{N}}k'}\ket{k'}^{\mathcal{O}}\\\nonumber
	=&\frac{1}{\mathcal{N}^2}\sum_{n=0}^{\mathcal{N}-1}\sum_{n'=0}^{\mathcal{N}-1}\sum_{k=0}^{\mathcal{N}-1}\sum_{k'=0}^{\mathcal{N}-1}\Big(e^{i\frac{2\pi m}{\mathcal{N}}k}\bra{k}^{\mathcal{O}}\ketbra{n}{n'}^\mathcal{O}e^{-i\frac{2\pi m}{\mathcal{N}}k'}\ket{k'}^{\mathcal{O}}\Big)\otimes \mathbb{L}_n\big\{\ast\big\}{\rho}_{\star}^{CW}\mathbb{L}_{n'}\big\{\ast\big\}^{\dagger}\\\nonumber
	=&\frac{1}{\mathcal{N}^2}\sum_{n=0}^{\mathcal{N}-1}\sum_{n'=0}^{\mathcal{N}-1}\sum_{k=0}^{\mathcal{N}-1}\sum_{k'=0}^{\mathcal{N}-1}\Big(e^{i\frac{2\pi m}{\mathcal{N}}(k-k')}\delta_{k,n}\delta_{n'k'}\Big) \mathbb{L}_n\big\{\ast\big\}{\rho}_{\star}^{CW}\mathbb{L}_{n'}\big\{\ast\big\}^{\dagger}\\\nonumber
	=&\frac{1}{\mathcal{N}^2}\sum_{k=0}^{\mathcal{N}-1}\sum_{k'=0}^{\mathcal{N}-1}\Big(e^{i\frac{2\pi m}{\mathcal{N}}(k-k')}\Big) \mathbb{L}_k\big\{\ast\big\}{\rho}_{\star}^{CW}\mathbb{L}_{k'}\big\{\ast\big\}^{\dagger}\\
	=&\Bigg[\frac{1}{\mathcal{N}}\sum_{k=0}^{\mathcal{N}-1}e^{i\frac{2\pi m}{\mathcal{N}}k}\mathbb{L}_k\big\{\ast\big\}\Bigg]{\rho}_{\star}^{CW}\Bigg[\frac{1}{\mathcal{N}}\sum_{k'=0}^{\mathcal{N}-1}e^{i\frac{2\pi m}{\mathcal{N}}k'}\mathbb{L}_{k'}\big\{\ast\big\}\Bigg]^{\dagger}.
\end{align}

As we have seen in Eq.~\eqref{eq:rho_cw_fmN_derv}, the phase terms $e^{i\frac{2\pi m}{\mathcal{N}}k}$ in front of $\mathbb{L}_k\big\{\ast\big\}$ is not relevant to the population terms in a corresponding density operator, or the walker's position probability. Therefore to prove Theorem 2, it is basically the same as what we did in the proof of Lemma 2. Then our goal is to prove that $\bra{\psi_i}\ket{\psi_j}=\delta_{i,j}$, where $\ket{\psi_i}=\bra{{\blacktriangleleft}}\ket{\Psi_i}$ with $\ket{\Psi_i}$ defined as $\mathbb{L}_{i}\big\{\ast\big\}(\ket{{\vartriangleright}}\otimes\ket{0})$.

If one notices that for $\forall~0\leq i \leq\mathcal{N}-1$, the relationship between $\mathbb{L}_{i}\big\{\ast\big\}$ and $\mathbb{L}_{i-1}\big\{\ast\big\}$ can be viewed as the result of applying a \emph{2}$-$Switch. Then our method in proving Lemma 2 also applies here, as well as the similar condition of $\sigma_i\geq\sigma_{i-1}$.
\section{Proof of Theorem 3}
In the special case of all $\sigma_i$ are equal, the population terms in $\varrho_{n|\phi^{\bot}_0}^W$ and $\varrho_{n-1|\phi^{\bot}_0}^W$ are related by
\begin{align}
	\label{eq:theorem_3_equality}\nonumber
	\bra{x}\varrho_{n|\phi^{\bot}_0}^W\ket{x}=\bra{x-2\sigma_i}&\varrho_{n-1|\phi^{\bot}_0}^W\ket{x-2\sigma_i},\\
	\text{for~}&\forall~n\geq 1.
\end{align}

By a straightforward calculation, one finds that the (non-normalized) distribution associated with $\bra{x}\varrho_{0|\phi^{\bot}_0}^W\ket{x}$ is 
\begin{align}
	\label{eq:two_step_distr}
	\text{Prob\{}x=i\}=
    \begin{cases}
      ~~1/4 & \text{if $i=-2\mathcal{N},-2\mathcal{N}+2$}\\
      ~~0 & \text{otherwise},
    \end{cases}       
\end{align}
if $\sigma_i=2$, and for another case, that is when $\sigma_i=4$, we have
\begin{align}
	\label{eq:four_step_distr}
	~\text{Prob\{}x=i\}=
    \begin{cases}
      ~~1/16 & \text{if $i=-4\mathcal{N},-4\mathcal{N}+2, -4\mathcal{N}+4, -4\mathcal{N}+6$}\\
      ~~0 & \text{otherwise}.
    \end{cases}       
\end{align}
Therefore using the property described by Eq.~\eqref{eq:theorem_3_equality}, one connects $\mathcal{N}$ shifted copies of the above distributions to obtain $\rho_{|\phi^{\bot}_0,\mathcal{F}_i^{\mathcal{N}}}^W$. Also, since all $\rho_{|\phi^{\bot}_0,\mathcal{F}_i^{\mathcal{N}}}^W$ are equivalent up to some phase differences in the coherence terms, we wrap all the $\mathcal{N}$ states together, and the probability for which is
\begin{align}
	\label{eq:theorem_prob_calc}\nonumber
	\sum_{i=0}^{\mathcal{N}-1}\text{Tr}\Big[\rho_{|\phi^{\bot}_0,\mathcal{F}_i^{\mathcal{N}}}^W\Big]=&\sum_{i=0}^{\mathcal{N}-1}\text{Tr}\Big[\frac{1}{\mathcal{N}^2}\sum_{n=0}^{\mathcal{N}-1}\varrho_{n|\phi^{\bot}_0}^W\Big]\\\nonumber
	=&\frac{1}{\mathcal{N}^2}\sum_{i=0}^{\mathcal{N}-1}\sum_{n=0}^{\mathcal{N}-1}\text{Tr}\Big[\varrho_{n|\phi^{\bot}_0}^W\Big]\\\nonumber
	=&\frac{1}{\mathcal{N}^2}\sum_{i=0}^{\mathcal{N}-1}\sum_{n=0}^{\mathcal{N}-1}\frac{1}{\sigma_i}\\
	=&\frac{1}{\sigma_i}.
\end{align}
Here, we have used $1/\sigma_i$ to represent the summation over all probability amplitudes in Eq.~\eqref{eq:two_step_distr} and Eq.~\eqref{eq:four_step_distr}.